%% file: ms.tex
\def\eiso{E_{\rm iso}}

\def\ep{E_{\rm peak}}
\def\eop{E^{\rm obs}_{\rm peak}}
\def\esp{E^{\rm src}_{\rm peak}}

\documentclass[12pt,preprint]{aastex}

\slugcomment{Not to appear in Nonlearned J., 45.}

\shorttitle{Swift X-Ray Flashes}
\shortauthors{Sakamoto et al.}

\begin{document}


\title{Global Properties of X-Ray Flashes and X-Ray-Rich Gamma-Ray Bursts Observed by {\it Swift}}


\author{T. Sakamoto\altaffilmark{1,2}, 
D. Hullinger\altaffilmark{9}, 
G. Sato\altaffilmark{1,7},
R. Yamazaki\altaffilmark{8},
L. Barbier\altaffilmark{1}, 
S. D. Barthelmy\altaffilmark{1}, 
J. R. Cummings\altaffilmark{1,3}, 
E. E. Fenimore\altaffilmark{4},
N. Gehrels\altaffilmark{1}, 
H. A. Krimm\altaffilmark{1,6}, 
D. Q. Lamb\altaffilmark{10}
C. B. Markwardt\altaffilmark{1,5},
J. P. Osborne\altaffilmark{11},
D. M. Palmer\altaffilmark{4},
A. M. Parsons\altaffilmark{1},
M. Stamatikos\altaffilmark{1,2}, 
J. Tueller\altaffilmark{1},
}

\altaffiltext{1}{NASA Goddard Space Flight Center, Greenbelt, MD 20771}
\altaffiltext{2}{Oak Ridge Associated Universities, P.O. Box 117, 
 Oak Ridge, TN 37831-0117}
\altaffiltext{3}{Joint Center for Astrophysics, University of Maryland, 
	Baltimore County, 1000 Hilltop Circle, Baltimore, MD 21250}
\altaffiltext{4}{Los Alamos National Laboratory, P.O. Box 1663, Los
Alamos, NM, 87545}
\altaffiltext{5}{Department of Physics, University of Maryland, 
	College Park, MD 20742}
\altaffiltext{6}{Universities Space Research Association, 10211 Wincopin 
	Circle, Suite 500, Columbia, MD 21044-3432} 
\altaffiltext{7}{Institute of Space and Astronautical Science, 
JAXA, Kanagawa 229-8510, Japan}
\altaffiltext{8}{Department of Physics, Hiroshima University, Higashi-Hiroshima,
        Hiroshima, 739-8526, Japan}
\altaffiltext{9}{Moxtek, Inc., 452 West 1260 North, Orem, UT  84057}
\altaffiltext{10}{Department of Astronomy and Astrophysics, University
of Chicago, Chicago, IL, 60637}
\altaffiltext{11}{Department of Physics and Astronomy, University of
Leicester, LE1, 7RH, UK}


\begin{abstract}

We describe and discuss the spectral and temporal characteristics of
the prompt emission and X-ray afterglow emission of X-ray flashes (XRFs) 
and X-ray-rich gamma-ray bursts (XRRs) 
detected and observed by {\it Swift} between December 2004 and
September 2006.  We compare these characteristics to a sample of 
conventional classical 
gamma-ray bursts (C-GRBs) observed during the same period.  We confirm the
correlation between $\eop$ and fluence noted by others and find further
evidence that XRFs, XRRs and C-GRBs form a continuum.  We also confirm that 
our known redshift sample 
is consistent with the correlation between the peak energy in the GRB
 rest frame ($\esp$) and 
the isotropic radiated energy ($\eiso$), so called the $\esp$-$\eiso$ 
relation.  The 
spectral properties of X-ray afterglows of XRFs and C-GRBs are similar, 
but the temporal properties of XRFs and C-GRBs are quite
different.  We found that the light curves of C-GRB afterglows show a break to 
steeper indices (shallow-to-steep break) at much earlier times than do 
XRF afterglows.  Moreover, the overall luminosity of XRF X-ray afterglows 
is systematically smaller by a factor of two or more compared to that of 
C-GRBs.  These distinct differences between the 
X-ray afterglows of XRFs and C-GRBs may be the key to understanding not only the 
mysterious shallow-to-steep break in X-ray afterglow light curves, but also the unique 
nature of XRFs.   

\end{abstract}




\section{Introduction}

Despite the rich gamma-ray burst (GRB) sample provided by BATSE 
\citep[e.g.,][]{paciesas1999,kaneko2006}, Beppo{\it SAX}
\citep[e.g.,][]{frontera2004}, $Konus$-$Wind$ \citep[e.g.,][]{ulanov2004}, 
and $HETE$-2 \citep[e.g.,][]{barraud2003,sakamoto}, 
the emission properties of GRBs 
are still far from being well-understood.  In recent years, however, another 
phenomenon that resembles GRBs 
in almost every way, except that the flux comes mostly 
from X rays instead of $\gamma$ rays, has 
been discovered and studied.  This new class of bursts has been 
dubbed ``X-ray flashes'' (XRFs; \citet{heise2,barraud2003,sakamoto}), 
and there is strong evidence
to suggest that ``classical'' GRBs (hereafter C-GRBs) and XRFs are 
closely-related phenomena. Understanding what physical processes lead 
to their differences could yield important insights into their nature 
and origin.

\citet{strohmayer} identified 22 bursts observed by 
$Ginga$ that occurred between March of 1987 and October of 1991, and for 
which the spectra could be reliably analyzed.  
About 36\% of GRBs observed by $Ginga$ had very soft spectra.  
They noted that these bursts resembled BATSE long GRBs in duration and 
general spectral shape, but the peak energies of the $\nu$F$_{\nu}$
spectrum, $\eop$, extended to lower values than those of the BATSE
bursts \citep{preece2000,kaneko2006}.  
\citet{heise2} reported that among the sources imaged by the Wide Field 
Cameras (WFCs) on board {\it Beppo}SAX was a class of fast
X-ray transients with durations less than 1000 s that 
were not ``triggered'' (that is, detected) by the Gamma Ray Burst
Monitor (GRBM).  This became their working definition of XRFs.  
\citet{kippen} searched for C-GRBs and XRFs which were observed 
simultaneously by WFC and BATSE.  They found 36 C-GRBs and 17 XRFs in a 
3.8-year period.  Joint WFC and BATSE spectral analysis was performed 
for the sample, and they found that XRFs have a significantly lower $\eop$ 
compared with C-GRBs.  They also found that there is no 
systematic difference between XRFs and C-GRBs in their low-energy 
photon indices, high-energy energy 
photon indices, or durations.  
The systematic spectral analysis of a sample of 45 {\it HETE-2} GRBs confirmed 
these spectral and temporal characteristics of XRFs.  
It is worth noting that nine out of sixteen XRF samples of 
{\it HETE-2} have $\eop$ $<$ 20 keV \citep{barraud2003,sakamoto}.  

Although the XRF prompt emission properties have been studied, 
until the launch of {\it Swift}~\citep{gehrels}, 
only a handful of X-ray afterglows associated with XRFs were reported.  
\citet{alessio2005} studied the prompt and afterglow emission of XRFs and 
X-ray-rich GRBs (XRRs) observed by {\it Beppo}SAX and {\it HETE-2}.  
They found that the XRF and XRR afterglow 
light curves seem to be similar to those of C-GRBs, including the break 
feature in the light curves.  
They also investigated the off-axis viewing scenarios of XRFs 
for the top-hat shaped jet \citep{yamazaki2002,yamazaki2004}, the universal 
power-law shaped jet \citep{rossi2002,zhang2002,lamb2005}, and the Gaussian jet \citep{zhang2004}, 
and concluded that these models might be consistent with the data.  
Their sample, however,  
only contains 9 XRFs/XRRs with measured X-ray afterglows.  Furthermore, 
the data points in the X-ray light curves were not well sampled, 
so that there are large uncertainties in the decay indices and the overall 
structures of the light curve in most cases.  Moreover, since the X-ray 
afterglow observations began $>$ 10$^{4}$ seconds after the trigger, 
their sample 
is able to say little about the early afterglow properties, which contain rich 
information that can constrain jet models for XRFs.  
Other XRF 
theoretical models are the inhomogeneous jet model \citep{toma2005}, the 
internal shock emission from high bulk Lorentz factor shells 
\citep{mochkovitch2003,barraud2005}, the external shock emission 
from low bulk Lorentz factor shells \citep{dermer1999,dermer2003}, 
and the X-ray emission from the hot cocoon of the GRB jet if viewed from 
off-axis \citep{meszaros2002,woosley2003}.  

Because of the sophisticated on-board localization capability of the {\it Swift} 
Burst Alert Telescope (BAT; \citet{barthelmy}) and the fast spacecraft 
pointing of {\it Swift}, more than 90\% of {\it Swift} GRBs have an 
X-ray afterglow observation from the {\it Swift} X-Ray Telescope (XRT;
\citet{burrows}) within a few hundred seconds after the trigger.  
Due to the fact that BAT is sensitive to relatively low energies 
(15-150 keV) and also a large effective area ($\sim$ 1000 cm$^{2}$ at 20 keV 
for a source on-axis), BAT is detecting also XRFs and XRRs.  However, 
because of the BAT's lack of response below 15 keV, it is very 
challenging to detect XRFs with $\eop$ of a few keV which 
dominated the XRF samples of the {\it Beppo}SAX and {\it HETE-2} 
\citep[e.g.,][]{kippen,sakamoto}.  Nonetheless, {\it Swift} has 
an unique capability for studying the detailed X-ray afterglow properties 
just after the burst for XRFs and XRRs with $\eop$ $\gtrsim$ 20 keV 
for the first time.  

The systematic study of the X-ray emissions of GRBs observed by XRT 
reveals a very complex power-law decay behavior consisting typically of 
an initial very 
steep decay (t$^{\alpha}$ with $-10 \lesssim \alpha_{1} \lesssim -2$) 
\citep[e.g.,][]{obrien2006,sakamoto2007}, followed by 
a shallow decay ($-1 \lesssim \alpha_{2} \lesssim 0$), followed by 
a steeper decay ($-2 \lesssim \alpha_{3} \lesssim -1$) 
\citep[e.g.,][]{nousek2006, obrien2006,willingale2007}, sometimes followed by 
a much steeper decay ($\alpha_{4} \lesssim -2$) 
\citep[e.g.,][]{willingale2007} and, in some cases (about 50\%), overlaid X-ray flares 
\citep[e.g.][]{burrows2005,chincarini2007,kocevski2007}.  Although there 
is increasing evidence that the initial very steep decay component 
$\alpha_{1}$ is a 
tail of the GRB prompt emission \citep[e.g.,][]{liang2006,sakamoto2007}, 
the origin of the phase from a shallow $\alpha_{2}$ to a steeper decay 
$\alpha_{3}$ (hereafter a shallow-to-steep decay) 
is still a mystery.  Moreover, not all GRBs have a 
shallow-to-steep decay phase in 
their X-ray afterglow light curves.  Thus, it is very important to investigate the 
X-ray afterglow light curves of bursts along with their prompt emission properties 
to find a difference in their characteristics between C-GRBs and XRFs.  

In this paper, we report the systematic study of the prompt and afterglow 
emission of 10 XRFs and 17 XRRs observed by {\it Swift} from December 2004 
through September 2006.  
Although the data from {\it Swift} BAT is the primary 
dataset for investigation 
of the prompt emission properties, we also use information 
from $Konus$-$Wind$ and {\it HETE-2} as reported on the Gamma-ray 
burst Coordinate
Network\footnote{http://gcn.gsfc.nasa.gov/gcn\_main.html} 
or in the literature, 
when available, to obtain better 
constraints on $\eop$.  We focus on X-ray afterglow properties observed by 
{\it Swift} XRT in this study.  In \S 2, we discuss our classification of GRBs, 
the analysis methods of the BAT and the 
XRT data, and the selection criteria of our sample.  
In \S 3 and \S4, we show the 
results of the prompt emission and the X-ray afterglow analysis, 
respectively.  We found distinct differences between XRFs and C-GRBs 
in the shape and in the overall luminosity of X-ray afterglows.  
We discuss the implications of our results in \S 5.  Our 
conclusions are summarized in \S 6.  
We used the cosmological parameters of $\Omega_{\rm m}$ = 0.3, 
$\Omega_{\rm \Lambda}$ = 0.7, and $H_{0}$ = 70 km s$^{-1}$ Mpc$^{-1}$.  
The quoted errors are at the 90\% confidence level unless we state otherwise.  

\section{Analysis}

\subsection{Working Definition of {\it Swift} GRBs and XRFs}

The precise working 
definitions adopted by others who have studied XRFs have tended
(understandably) 
to be based on the characteristics and energy sensitivities of the 
instruments that collected the data
\citep{gotthelf,strohmayer,heise2,sakamoto}.  The effective area of 
the BAT is sufficiently different from these other instruments that 
none of the definitions previously adopted are quite suitable 
\citep{band2003,band2006}.  We desire a definition, however, 
that will correspond to previous definitions so that we may reliably 
compare the characteristics of the
BAT-detected XRF population with those from other missions.
\citet{sakamoto} defined XRFs in terms of the fluence
ratio 
$S_{X}$(2 -- 30 keV)/$S_{\gamma}$(30 -- 400 keV) and 
C-GRBs, XRRs, and XRFs were classified according to
this fluence ratio.  \citet{sakamoto} noted a strong 
correlation between the observed spectral peak energy $\eop$
and the fluence ratio.  They found that the border $\eop$
between XRFs and XRRs is $\approx 30$ keV, and the
border $\eop$ between XRRs and C-GRBs is $\approx 100$ keV.

In the BAT energy range, a fluence ratio of
$S$(25 -- 50 keV)/$S$(50 -- 100 keV) is more natural and easier 
to measure with
confidence.  We therefore chose our working definition in terms of this
ratio.  In order to ensure that our definition is close to that adopted
by \citet{sakamoto}, we calculated the fluence ratio of a burst for which
the parameters of the Band function\footnote{$f(E) = K_{1} 
E^{\Gamma_{1}} \exp[-E(2+\Gamma_{1})/\ep]$ if 
$E < (\Gamma_{1} - \Gamma_{2}) \ep/(2+\Gamma_{1})$ and 
$f(E) = K_{2} E^{\Gamma_{2}}$ if $E \geq 
(\Gamma_{1} - \Gamma_{2}) \ep/(2+\Gamma_{1})$.} \citep{band1993}
are $\Gamma_{1}=-1$, 
$\Gamma_{2}=-2.5$, and
$\eop=30$ keV.  These values of $\Gamma_{1}$ and of $\Gamma_{2}$ are typical
of the distributions for XRFs, XRRs, and C-GRBs found by 
BATSE \citep{preece2000,kaneko2006}, {\it Beppo}SAX \citep{kippen} 
and {\it HETE-2} \citep{sakamoto}.  
The ratio thus found is 1.32.
We likewise calculated the fluence ratio of a burst for which
$\Gamma_{1}=-1$, $\Gamma_{2}=-2.5$, and
$\eop=100$ keV, which was found to be 0.72.  Our working definition
of XRFs, XRRs, and C-GRBs thus becomes:

\begin{eqnarray}
 S(\textrm{25 -- 50 keV})/S(\textrm{50 -- 100 keV}) &\le 0.72
& ~{\rm C-GRB} \nonumber\\
 0.72 < S(\textrm{25 -- 50 keV})/S(\textrm{50 -- 100 keV}) &\le 1.32
& ~{\rm XRR}\\
 S(\textrm{25 -- 50 keV})/S(\textrm{50 -- 100 keV}) &> 1.32
& ~{\rm XRF} \nonumber
\end{eqnarray}

\begin{figure}
\includegraphics[angle=-90,scale=.65]{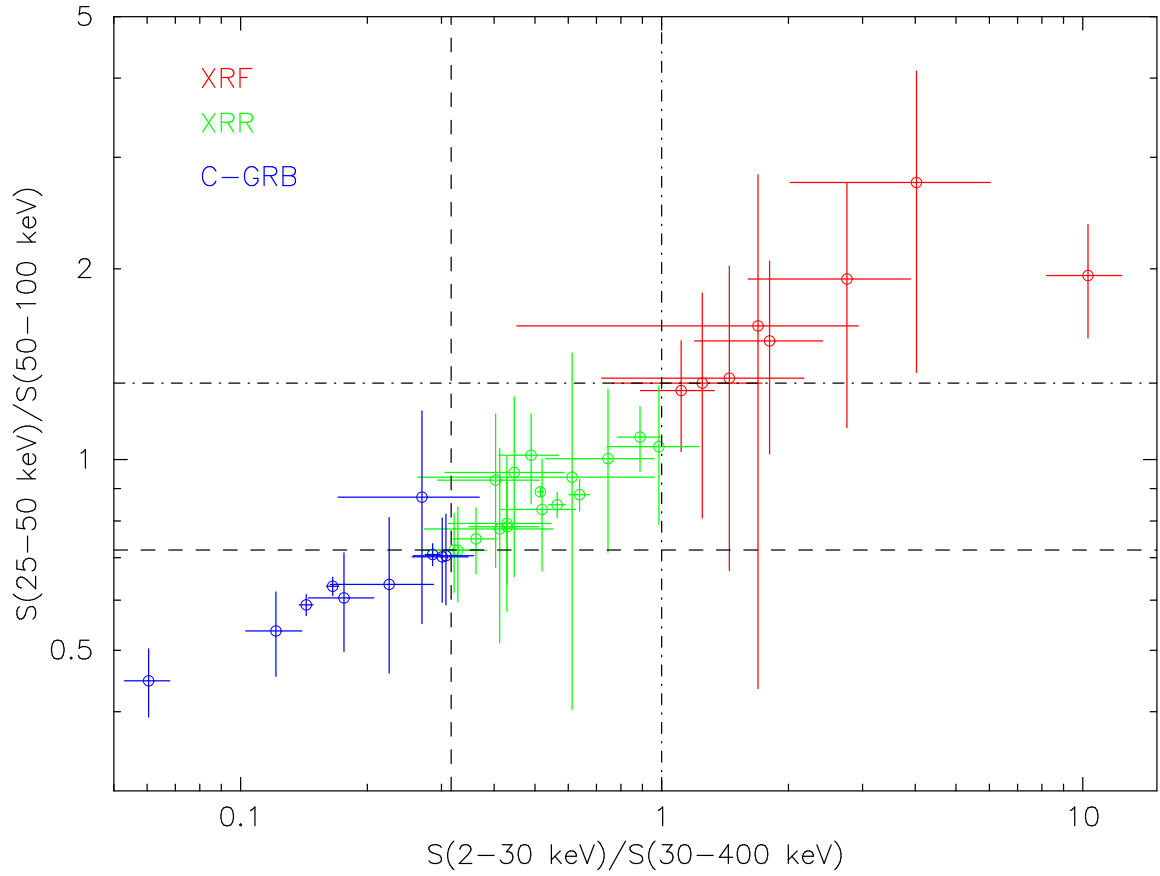}
\caption{$S$(2--30 keV)/$S$(30--400 keV) and 
$S$(25--50 keV)/$S$(50--100 keV) 
fluence ratios of {\it HETE-2} bursts.  The dashed and 
dash-dotted lines correspond to the borders between C-GRBs and XRRs, 
and between XRRs and XRFs, respectively.  
\label{here_fluence_ratio}}
\end{figure}

To check the consistency of our definition, we calculated $S(\textrm{25 -- 50 keV})$ 
and $S(\textrm{50 -- 100 keV})$ of the {\it HETE-2} sample using the best fit 
time-averaged spectral parameters reported on \citet{sakamoto}.  The 90\% error in 
the fluences is calculated by scaling the associated error in the normalization of the 
best fit spectral model.  As shown in Figure \ref{here_fluence_ratio}, our definition 
is consistent with the {\it HETE-2} definition of XRFs, XRRs, and C-GRBs \citep{sakamoto}.  

\subsection{{\it Swift} BAT Data Analysis}

All the event data from {\it Swift} BAT is available through HEASARC at 
Goddard Space Flight Center.  We used the standard BAT software 
(HEADAS 6.1.1) and the latest 
calibration database (CALDB: 2006-05-30).  The burst pipeline script, 
{\tt batgrbproduct}, was used to process the BAT event data.  
The {\tt xspec} spectral fitting tool (version 11.3.2) was used to fit each 
spectrum.   

For the time-averaged spectral analysis, we use the time interval from
0\% to 100\% of the total burst fluence ($t_{100}$ interval) calculated by 
{\tt battblocks}.  
Since the BAT energy response generator, {\tt batdrmgen}, performs the 
calculation for a fixed single incident angle of the source, it will be 
a problem if the 
position of the source is moving during the time interval selected for 
the spectral 
analysis due to the spacecraft slew.  In this situation, we created the
response matrices for each five second period during the time interval 
taking into account the position
of the GRB in detector coordinates.  We then weighted these response
matrices by the five second count rates and created the averaged response 
matrices using {\tt addrmf}.  
Since the spacecraft slews about one degree per second in response to a GRB
trigger, we chose five second intervals to calculate
the energy response for every five degrees.

We fit each spectrum with a power-law (PL) model\footnote{$f(E) = K_{50} 
(E/50 {\rm keV})^{\Gamma}$ where $K_{50}$ is the normalization at 50 keV 
in units of photons cm$^{-2}$ s$^{-1}$ keV$^{-1}$.}
and a cutoff power-law (CPL) model\footnote{$f(E) = K_{50}(E/{\rm 50 keV})^{\Gamma} 
\exp(-E(2+\Gamma)/\ep)$.}.  
The best fit spectral model is determined 
based on the difference in $\chi^{2}$ between a PL and a CPL fit.  If 
$\Delta$ $\chi^{2}$ between a PL and a CPL fit is greater than 6 ($\Delta \chi^{2}
\equiv \chi^{2}_{\rm PL} - \chi^{2}_{\rm CPL}$ $>$ 6), we determine that a 
CPL model is a better 
representative spectral model for the data.  To quantify the significance of this 
improvement, we performed 10,000 spectral simulations taking into account the 
distributions of the power-law photon index in a PL fit, the fluence in the 15-150 keV band 
in a PL fit and the $t_{100}$ duration of the BAT GRBs 
\citep[e.g.,][]{bat1catalog}, and determined
how many cases a CPL fit gives 
$\chi^{2}$ improvements of equal or greater than 6 over a PL fit.  
We used the best fit normal distribution for the power-law photon index 
centering on 1.65 with $\sigma$ of 0.36.  
The best fit log-normal distribution is used for the fluence centering 
on S(15--150 keV) = $10^{-5.92}$ erg cm$^{-2}$  with $\sigma$ 
of S(15--150 keV) = $10^{0.59}$ erg cm$^{-2}$.  Also, the best fit log-normal 
distribution is used for the $t_{100}$ duration centering on 
$t_{100}$ = 10$^{1.74}$ s with $\sigma$ of $t_{100}$ = $10^{0.53}$ s.  
The BAT energy response matrix used in the simulation corresponds to an incident 
angle of 30$^{\circ}$ which is the average of the BAT GRB samples.  
We found equal or higher 
improvements in $\chi^{2}$ in 62 simulated spectra out of 10,000.  Thus, the chance 
probability of having an equal or higher $\Delta \chi^{2}$ of 6 with a CPL model when 
the parent distribution is a case of a PL model is 0.62\%.  

\begin{figure}
\includegraphics[angle=90,scale=-0.65]{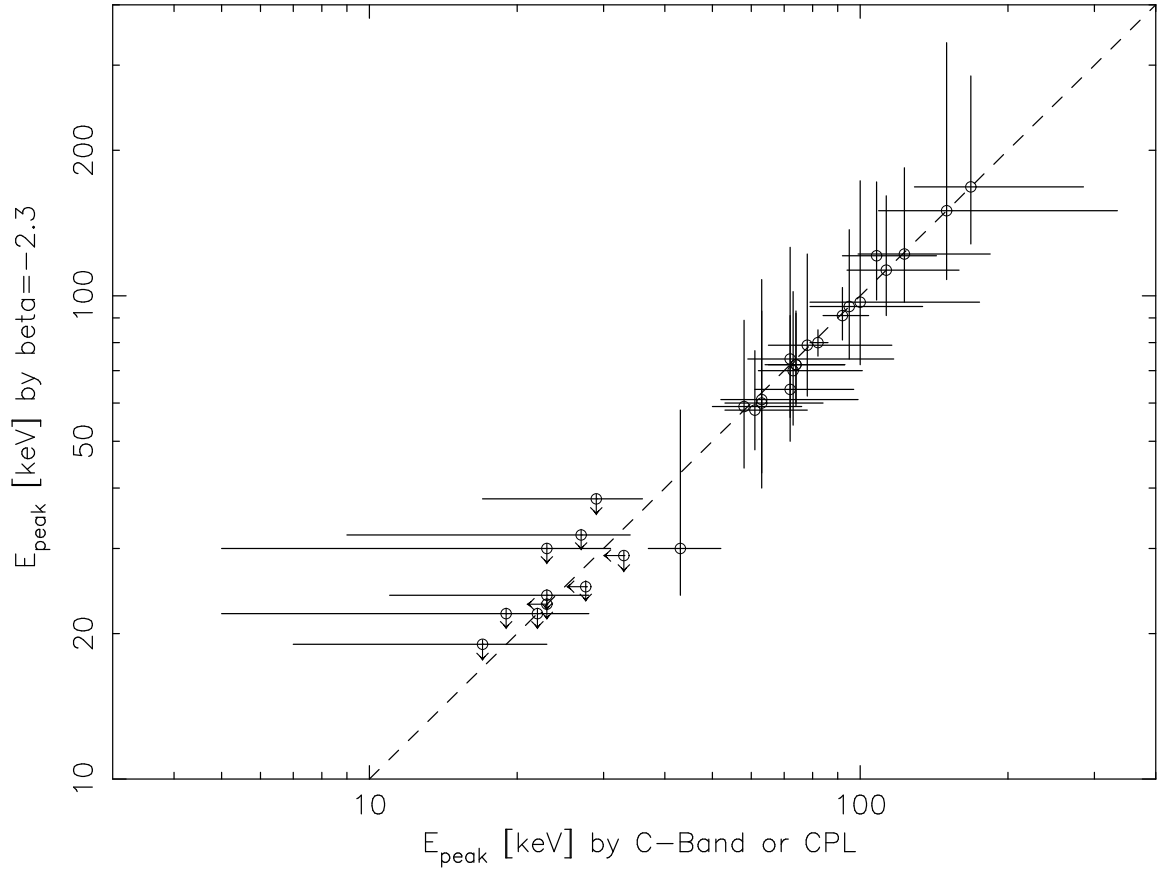}
\caption{The relationship between $\eop$ derived by the Band function with 
a fixed high-energy photon index $\Gamma_{2} = -2.3$ and $\eop$ derived by the 
C-Band function or a CPL model.}
\label{ep_bat_ep_betan23}
\end{figure}

Because of the narrow energy band of the BAT, most of the $\eop$ measured 
from the BAT spectral data are based on a CPL fit, but not on the Band 
function fit.  For XRFs, we apply a 
{\it constrained} Band (C-Band) function method \citep{sakamoto2} to constrain $\eop$.  
However, there is a systematic problem in the $\eop$ values derived by different 
spectral models.  In particular, for the bursts for which the true spectral shape is the 
Band function, there is a known effect that $\eop$ derived from 
a CPL model fit has a systematically higher value than $\eop$ derived from a Band function 
fit \citep[e.g.,][]{kaneko2006,cabrera2007}.  
To investigate this effect, we fit all the BAT GRB spectra for which $\eop$ are derived only 
from the BAT data with a Band function with the high-energy photon index 
fixed at $\Gamma_{2}$ to $-2.3$.  Figure
\ref{ep_bat_ep_betan23} shows $\eop$ derived by the Band function fixing
$\Gamma_{2}$ = $-2.3$ and $\eop$ derived by a CPL or a C-Band function.  
The $\eop$ values derived by the Band function with fixing 
$\Gamma_{2}$ = $-2.3$ and by a CPL model agree within errors.  
Most of $\eop$ values derived by a C-Band function also agree with $\eop$ derived by the 
Band function with fixing $\Gamma_{2}$ = $-2.3$ to within errors.  
Therefore, we conclude that the systematic error in $\eop$ derived by 
different spectral models is negligible compared to that of the statistical 
error assigned to 
$\eop$ derived from the BAT spectral data alone.  Note that the BAT spectral data 
include the systematic errors which are introduced to reproduce the canonical 
spectrum of the Crab nebula observed at various incident angles \citep{bat1catalog}.  

To perform the systematic study using 
the BAT data, we only selected bursts for which 
the full BAT event data are 
available\footnote{We exclude bursts such as GRB 050820A, GRB
051008, and GRB 060218 because of incomplete event data.}.  

\subsection{{\it Swift} XRT Data}

We constructed a pipeline script to perform the XRT analysis in a 
systematic way.  This pipeline script analysis is composed of four parts: 
1) data download from the {\it Swift} Science Data Center (SDC), 2) an image 
analysis to find the source (X-ray afterglow) and background regions, 
3) a temporal analysis to construct and fit the light curve, and 
4) a spectral analysis.  The screened event data of the Window Timing 
(WT) mode and the Photon Counting (PC) mode are downloaded from the SDC and 
used in our pipeline process.  For the WT mode, only the data 
of the first segment number (001) are selected.  All available PC mode data are applied.  
The standard grades, grades 0-2 for the WT mode and 0-12 for
the PC mode, are used in the analysis.  
The analysis is performed in the 0.3--10 keV band.  
The detection of an X-ray afterglow is done automatically using {\tt ximage} 
assuming that an afterglow is the 
brightest X-ray source located within 4$^{\prime}$ from the BAT on-board position.   
However, in cases where a steady 
cataloged bright X-ray source is misidentified as an afterglow, we specify 
the coordinates of 
the X-ray afterglow manually.  The source region of the PC mode is 
selected as a circle of 47$^{\prime\prime}$ radius.  The background region 
of the PC mode is an annulus in an outer radius of 150$^{\prime\prime}$ 
and an inner radius 
of 70$^{\prime\prime}$ excluding the background X-ray sources 
detected by {\tt ximage} in circular regions of 47$^{\prime\prime}$ radius.  
For the WT data, the rectangular region of $1.6^{\prime} \times 6.7^{\prime}$ 
is selected as a foreground region using an afterglow position derived 
from the PC mode data as the center of the region.  The background region is 
selected to be a square region of $6.7^{\prime}$ on a side excluding 
a $2.3^{\prime} \times 6.7^{\prime}$ rectangular region centered at the 
afterglow position.  The light curve is binned based on the number of 
photons required to meet at least 5$\sigma$ for 
the PC mode and 10$\sigma$ for the WT mode in each light curve bin.  
The light curve fitting starts with a single power-law.  Then, additional 
power-law components are added to minimize $\chi^{2}$ of the fit.  
Complicated structures such as X-ray flares are also well fitted with 
this algorithm.  
Although our pipeline script fits the XRT light curve automatically 
for every GRB trigger by this algorithm, we excluded the time intervals  
during the X-ray flares from the light curve data by visual inspections 
before doing the fit by our method because the 
understanding of the overall shape of the light curve is the primary 
interest in our study.  
The ancillary response function (ARF) files 
are created by {\tt xrtmkarf} for the WT and the PC mode data individually.  
The spectral fitting is performed by {\tt xspec 11.3.2} using an absorbed 
power-law model\footnote{$wabs*wabs*pegpwrlw$ or $wabs*zwabs*pegpwrlw$ model 
in xspec} for both the WT and the PC mode data.  
For an absorption model, we fix the galactic absorption of 
\citet{dickey1990} at the GRB location, and then, add an additional 
absorption to the model.  We use {\tt xspec} 
$zwabs$ model for known redshift GRBs applying the measured redshift to calculate 
the absorption associated to the source frame of GRBs.  
The spectra are binned to at least 20 counts in each 
spectral bin by {\tt grppha}.  The conversion factor from a count rate 
to an unabsorbed 0.3--10 keV energy flux is also calculated based on
the result of the time-integrated spectral analysis.   

A ``pile-up'' correction \citep[e.g.,][]{romano2006,nousek2006,evans2007} 
is applied during our pipeline process.  It assumes a 
``pile-up'' effect exists whenever the uncorrected count rate in the 
processed light curve exceeds 0.6 counts/s and 100 counts/s for the PC and the WT 
modes respectively.  Only the time intervals which are affected by the 
``pile-up'' as described in our definition above have corrections 
applied.  Although the area of the spectral region affected by pile-up depends 
on its count rate, the script always eliminates a central area within 
7$^{\prime\prime}$ radius for the PC data and a 14$^{\prime\prime} 
\times 6.7^{\prime}$ box region for the WT data.  The count rate derived 
from the region excluding the central part is corrected by taking into 
account the shape of the ARF at an averaged photon energy.  The spectral 
analysis is performed using only the data of $<$ 0.6 counts/s for the the 
PC mode and $<$ 100 counts/s for the WT mode.  

Two GRBs in our sample, GRB 050713A and GRB 060206 have a background X-ray 
source $\sim 25^{\prime\prime}$ and 
$\sim 10^{\prime\prime}$, respectively, from the position of the afterglow.  
Since it is difficult to exclude the contamination from the very closely located 
background source, we excluded the last 
portion of the light curves which have a flattening that is 
very likely due to the contamination from the background source.  

\subsection{Sample of GRBs}

\begin{figure}
\includegraphics[angle=-90,scale=.65]{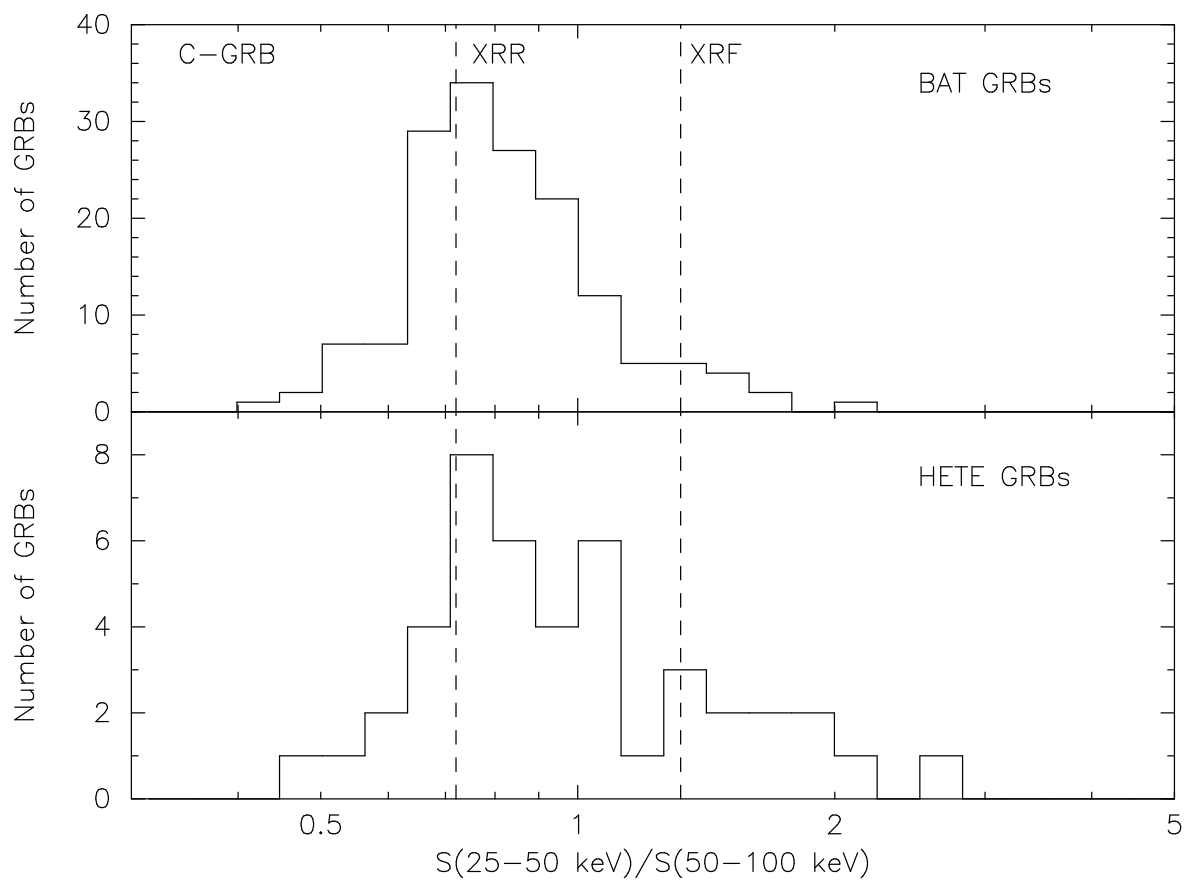}
\caption{Distributions of the fluence ratio 
$S(25-50~{\rm keV})/S(50-100~{\rm keV})$ for the BAT (top) 
and the {\it HETE-2} (bottom).  The dashed lines corresponds to 
the borders between C-GRBs and XRRs, and between XRRs and XRFs.}
\label{bat_hete_s23_hist}
\end{figure}

We calculated the fluence ratio between the 25--50 keV and the 50--100 keV
bands derived from a PL model using the BAT time-averaged 
spectrum for all 
{\it Swift} bursts detected between December 2004 and September 2006.  Then we 
classified these GRBs using the definition described in \S 2.1.  
Out of a total of 158 long GRBs, we classified 10 as XRFs, 97 as XRRs, 
and 51 as C-GRBs.  
The distribution of the fluence ratio S(25--50 keV)/S(50--100 keV) for the 
158 long GRBs is shown in Figure \ref{bat_hete_s23_hist}.  Similar to the 
{\it HETE-2} results \citep{sakamoto}, the figure clearly shows that {\it
Swift's} XRFs, XRRs, and C-GRBs also form a single broad distribution.  
This figure also clearly shows that the ratio of the number of BAT XRFs 
to BAT XRRs is smaller than that of the {\it HETE-2} XRF samples.  
As discussed in \citet{band2006}, the numbers of each GRB class strongly 
depend on the sensitivity of the instrument.  This problem becomes more 
serious for the instruments which do not cover a wide energy range,  
such as the BAT.  Thus, we will not discuss the absolute numbers of 
each GRB class in this paper.  

Since the determination of $\eop$ is crucial for our study, we only 
select GRBs having values for $\eop$ that can be determined from 
the BAT data alone or from using the data from other GRB 
instruments ($Konus$-$Wind$ and {\it HETE-2}).  
Since we can use the C-Band function method for XRFs to constrain 
$\eop$ if the photon index $\Gamma$ in 
a PL fit is much steeper than $-2$ in the BAT spectrum, we select all 
bursts which have $\Gamma < -2$ at a 90\% confidence level.  
We exclude GRB 041224 from our sample because there is no XRT observation.  
We also exclude GRB 060614 because of no report on the time-averaged 
spectral parameters by $Konus$-$Wind$ \citep{golenetskii2006b}.  
Based on these selection criteria, a total of 41 GRBs are selected,   
including 10 XRFs, 17 XRRs, and 14 C-GRBs.  

\section{Prompt Emission}

\begin{figure}
\includegraphics[angle=-90,scale=.65]{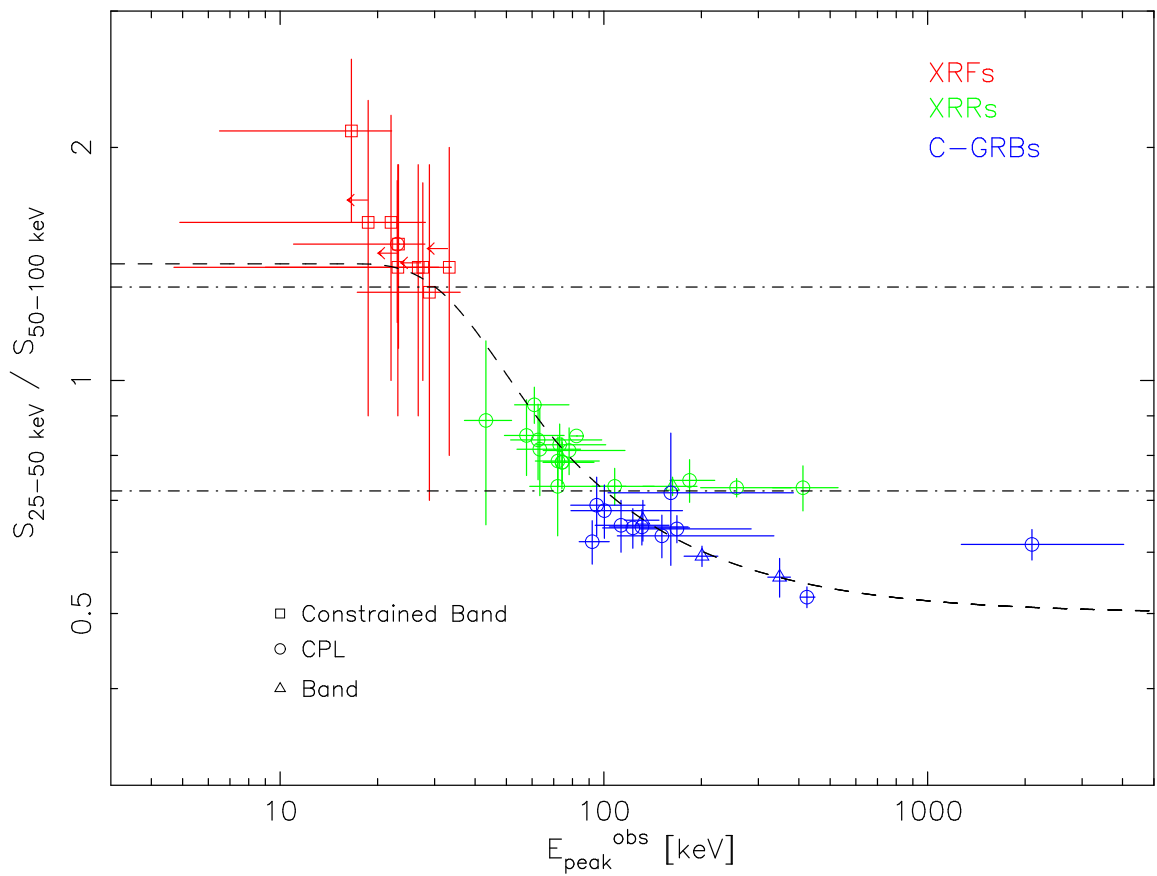}
\caption{$S$(25 -- 50 keV)/$S$(50 -- 100 keV)
fluence ratios and $\eop$ values
of BAT-detected bursts.  The dashed line shows the fluence 
ratios as a function of $\eop$ assuming $\Gamma_{1} =-1$ and 
$\Gamma_{2}=-2.5$ in the Band function.  
The dash-dotted lines indicate 
the boundaries between XRFs, XRRs, and C-GRBs (equation (1) in the text).  
\label{sr_ep}}
\end{figure}

The spectral properties of the prompt emission for our 41 GRBs are summarized 
in Table \ref{tbl1}.  Figure~\ref{sr_ep} shows the 
$S$(25 -- 50 keV)/$S$(50 -- 100 keV) fluence ratio verses $\eop$.  
As seen in the figure, $\eop$ of the BAT GRBs ranges from a few tens of 
keV to a few hundreds of keV.  This broad continuous distribution of $\eop$ is 
consistent with the {\it Beppo}SAX \citep{kippen} and the 
{\it HETE-2} \citep{barraud2003,sakamoto} results.  
The BAT GRBs follow well on the curve calculated assuming 
$\Gamma_{1}$ = $-1$ and $\Gamma_{2}$ = $-2.5$ for the Band function.  
The gap in the $S$(25 -- 50 keV)/$S$(50 -- 100 keV) 
fluence ratio from 0.8 to 1.2 in our sample is likely due to a 
selection effect.  Essentially, we selected bursts based on the measurement 
of $\eop$ for XRRs and C-GRBs.  This criterion is more or less 
equivalent to selecting the bursts based on their brightness.  
On the other hand, most of the XRFs were selected based on the photon index value 
in a PL fit ($\Gamma < -2$).  This is equivalent to selecting by the softness 
of the bursts.  Therefore, there is a different way to distinguish 
between XRFs, and XRRs and C-GRBs.  
Evidently, as shown in figure \ref{bat_hete_s23_hist}, there is no such gap 
in the histogram of the fluence ratios for the BAT GRBs if the whole 
burst sample has been examined.  Therefore, we believe that the 
gap in the fluence ratio at 0.8--1.2 is due to the way in which we selected 
the bursts.  

\begin{figure}
\includegraphics[angle=-90,scale=.65]{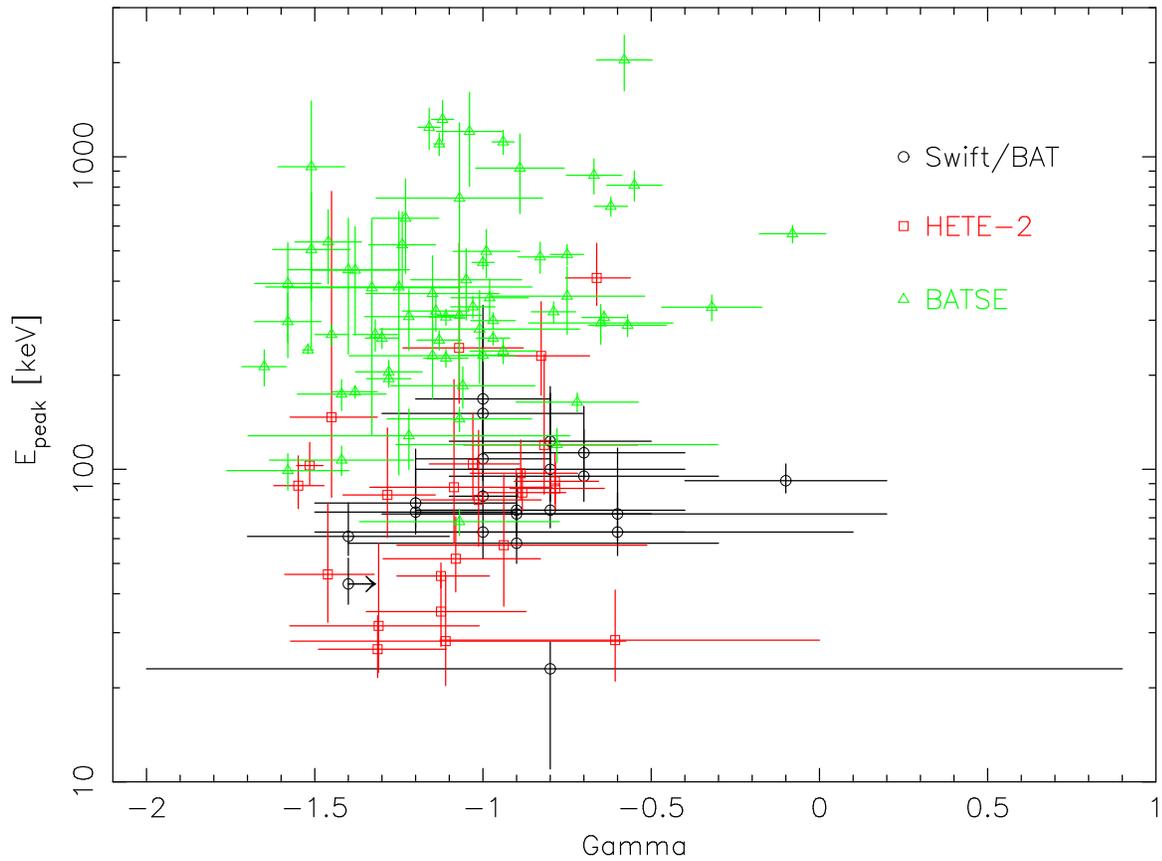}
\caption{Distribution of the low energy photon index $\Gamma$ and $\eop$ 
in a CPL model.  
The samples of BAT, {\it HETE-2}, and BATSE are shown in black circles, 
red squares, and green triangles, respectively.  \label{alpha_ep}}
\end{figure}

In Figure~\ref{alpha_ep}, we compare $\eop$ in a CPL fit and the low energy photon 
indices $\Gamma$ for the BAT, the {\it HETE-2} and the BATSE samples.  
For both the {\it HETE-2} \citep{sakamoto} and the BATSE \citep{kaneko2006} samples, 
we only plotted GRBs with a CPL model as the best representative model 
for the time-averaged spectrum to reduce the systematic differences in both 
$\Gamma$ and $\eop$ due to the different choices of spectral models 
\citep{kaneko2006}.  As seen in the figure, the range of $\Gamma$ 
values derived from the BAT data alone are consistent with 
the {\it HETE-2} and the BATSE results.  In addition, we have confirmed that 
the $\Gamma$ values for XRFs and XRRs (GRBs with $\eop$ $<$ 100 keV) 
cover the same range as for C-GRBs (GRBs with $\eop$ $>$ 100 keV) 
\citep{sakamoto}.  

\begin{figure}
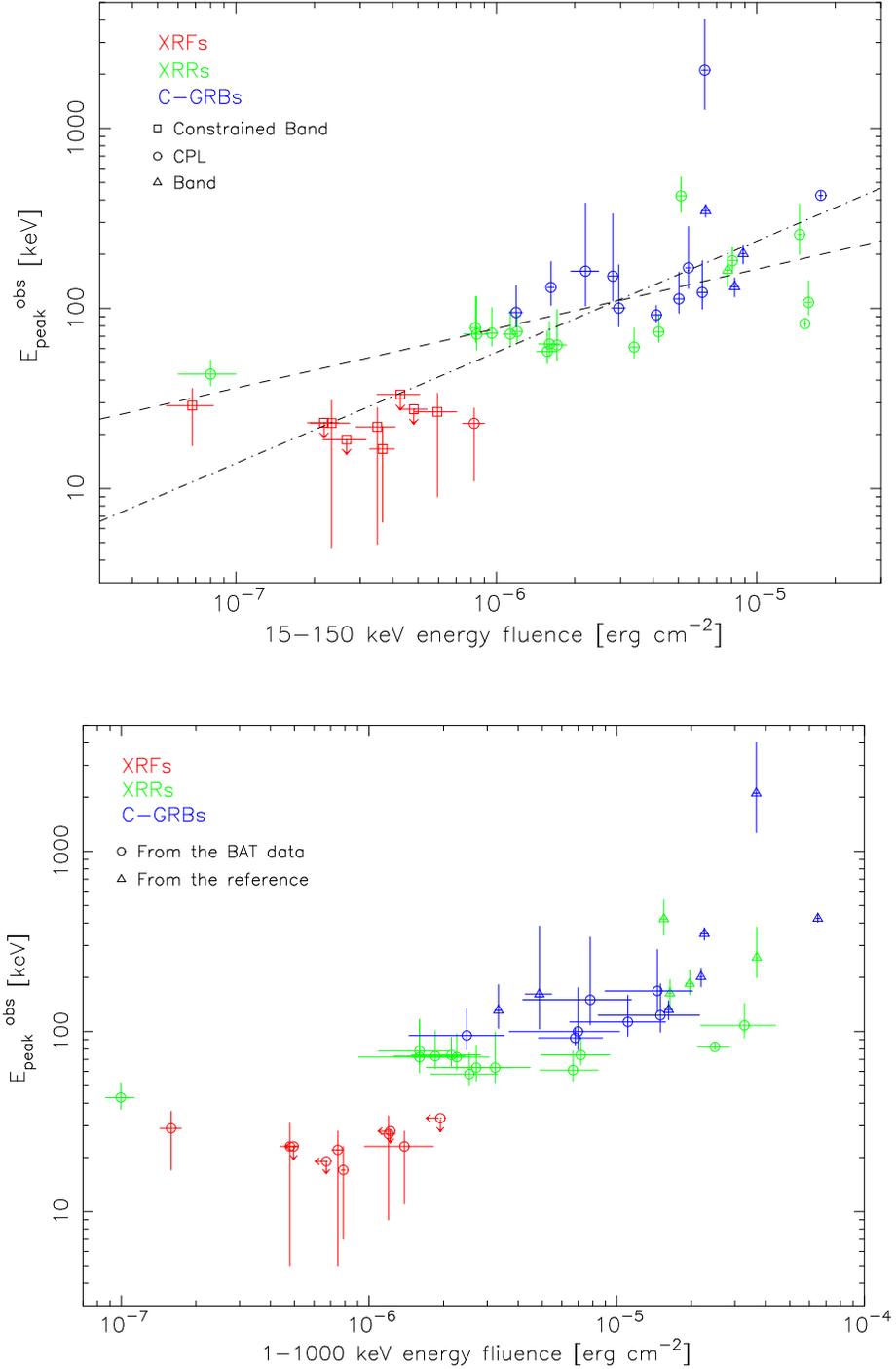

\centerline{
\includegraphics[angle=-90,scale=.5]{f6a.eps}}
\vspace{1cm}
\centerline{
\includegraphics[angle=-90,scale=.5]{f6b.eps}}
\caption{A plot of the 15 -- 150 keV fluence and peak spectral energy $\eop$
of XRFs (red), XRRs (green), and C-GRBs (blue) detected by BAT.  
The dashed and dash-dotted lines are the best fit 
to the data with and without taking into account the errors, 
and are given by 
$\log(\eop) = 3.87_{-0.16}^{+0.33} + (0.33 \pm 0.13) \log(S(15-150~{\rm keV}))$ and 
$\log(\eop) = (5.46 \pm 0.80) + (0.62 \pm 0.14) \log(S(15-150~{\rm keV}))$.  
Those bursts for which $\eop$ is derived from 
a $constrained$ Band function, a CPL, and the Band function are marked 
as squares, circles, and triangles, respectively.  \label{ep_stot}}
\end{figure}

The top panel of Figure~\ref{ep_stot} shows $\eop$ and the 15 -- 150 keV
fluence, S(15--150 keV), for the BAT GRBs.  
We note a correlation between $\eop$ and S(15--150 keV).  
For the purpose of the correlation study, we assigned the 
median of the 90\% confidence interval to be the 
best fit value of $\eop$, so that the errors would be symmetric.  
For cases in which we only have upper limits for $\eop$, 
we assigned the best fit values of $\eop$ to be the median of 0 
and 90\% upper limit, and we assigned the symmetric error to be 
half that value.  The linear correlation coefficient 
between log[S(15 - 150 keV)] and log($\eop$) is +0.76 for a sample 
of 41 GRBs using the best fit values.  
The best fit functions with and without 
taking into account the errors are $\log(\eop) = 3.87_{-0.16}^{+0.33} 
+ (0.33 \pm 0.03) \log[S(15 - 150~{\rm keV})]$ and $\log(\eop) = (5.46 \pm 0.80)
+ (0.62 \pm 0.14) \log[S(15 - 150~{\rm keV})]$, respectively.  

Since the fluence in the 15 -- 150 keV band is not a good quantity 
to examine the correlation with $\eop$ because of its narrow energy 
range of integration, we also investigate the correlation between $\eop$ 
and the fluence in the 1 -- 1000 keV band, S(1--1000 keV).  For GRBs which 
have the measurement of $\eop$ by the BAT data alone, we calculate S(1--1000 keV)  
directly from a spectral fitting process using the Band function.  Therefore, 
uncertainty in the spectral parameters in the Band function, especially in the 
high-energy photon index $\Gamma_{2}$ is also taken into account in an error 
calculation of the fluence.  For GRBs for which we use $\ep$ from the literature, 
we calculated the fluence using the spectral parameters presented in the 
literature, and the error associated in the normalization of the 
best fit spectral model is used to calculate an error of the fluence.  
If the reported best fit model is a CPL for these GRBs, we use $\Gamma_{2} = -2.3$ to 
calculate the fluence in the Band function.  The bottom 
panel of Figure~\ref{ep_stot} shows the distribution between $\eop$ and S(1--1000 keV).  

To take into account the errors 
associated with $\eop$ and S(1--1000 keV) in our calculation of the 
correlation coefficient, we generate 10,000 random 
numbers assuming a Gaussian distribution in $\eop$ and S(1--1000 keV) 
of the central value and the error for each GRB in the sample.  
For GRBs only having the upper limits in $\eop$ and/or S(1--1000 keV), 
we use an uniform distribution to generate the random numbers.  
Then, we calculate the linear correlation 
coefficient for the 10,000 burst sample in log[$\eop$]-log[S(1--1000 keV)] space, 
and make a histogram of the calculated correlation coefficient.   The highest 
peak and 68\% points from the highest value of the histogram are 
assigned as the central value and 1$\sigma$ interval of the correlation coefficient.  
We investigate the correlation 
coefficient for 1) GRBs with $\eop$ from a CPL model (sample A;
total 32 GRBs), 2) GRBs with a constrained $\eop$ from a C-Band 
model and a CPL model (sample B; total 37 GRBs), and 3) all 41 GRBs 
(sample C) to evaluate the systematic
effect of $\eop$ due to the different spectral models (C-Band 
vs. CPL).  The calculated correlation coefficients are 
$+0.50_{-0.12}^{+0.11}$, $+0.63_{-0.10}^{+0.08}$, and $+0.68_{-0.08}^{+0.07}$ 
(all in 1$\sigma$ error) for samples A, B and C respectively.  
The probabilities of such a correlation occurring by chance in each 
sample size are $3.4 \times 10^{-2}$ -- $2.4 \times 10^{-4}$, 
$5.8 \times 10^{-4}$ -- $5.3 \times 10^{-7}$, and $4.1 \times 10^{-5}$ -- 
$2.3 \times 10^{-8}$ in the 1 $\sigma$ interval for samples A, B 
and C respectively.  Thus, the correlation between $\eop$ and the
fluence is still significant even if we use the fluence in 
the 1--1000 keV band, and also take into account the $\eop$ derived by the different 
spectral models.  

\begin{figure}
\includegraphics[angle=-90,scale=.65]{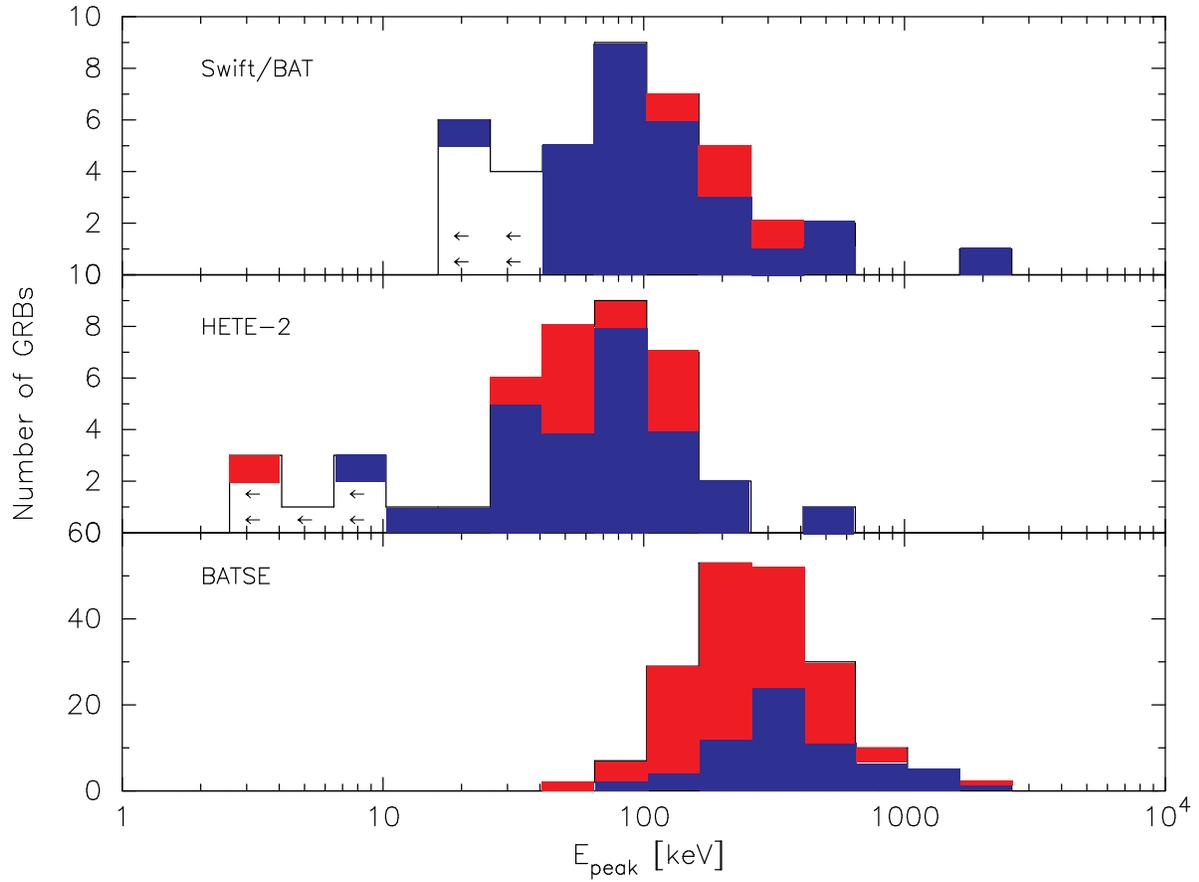}
\caption{Distribution of $\ep$ for the {\it Swift}/BAT, the {\it HETE-2}, and the 
BATSE samples.  The white, blue and red histograms are $\ep$ derived 
by the {\it constrained} Band function, a CPL model, and the Band 
function respectively.  The left side arrows are $\ep$ with upper limits.}
\label{ep_hist}
\end{figure}

The histograms of $\eop$ for the {\it Swift}/BAT, the {\it HETE-2} \citep{sakamoto} 
and the BATSE \citep{kaneko2006} samples are shown in Figure~\ref{ep_hist}.  
We notice a difference in the distributions of $\eop$ for the three GRB instruments, 
especially between the BAT (or the {\it HETE-2}) and the BATSE distributions.  
Applying the two-sample Kolmogorov-Smirnov (K-S) test to the $\eop$ 
distributions for the BAT and the {\it HETE-2} samples, the BAT and the BATSE samples, 
and the {\it HETE-2} and the BATSE samples, we find K-S test probabilities 
of 0.44, $2.3 \times 10^{-9}$, 
and $4.1 \times 10^{-16}$ respectively.  Based on these tests, we may conclude 
that the BATSE GRB samples have a systematically higher $\eop$ than the BAT and 
the {\it HETE-2} samples.  This is probably because not only the BATSE energy range is 
higher than those other instruments but also the current BATSE spectral 
catalog only contains the bright GRBs, therefore systematically selecting higher 
$\eop$ GRBs in the catalog \citep{kaneko2006}.  

\begin{figure}
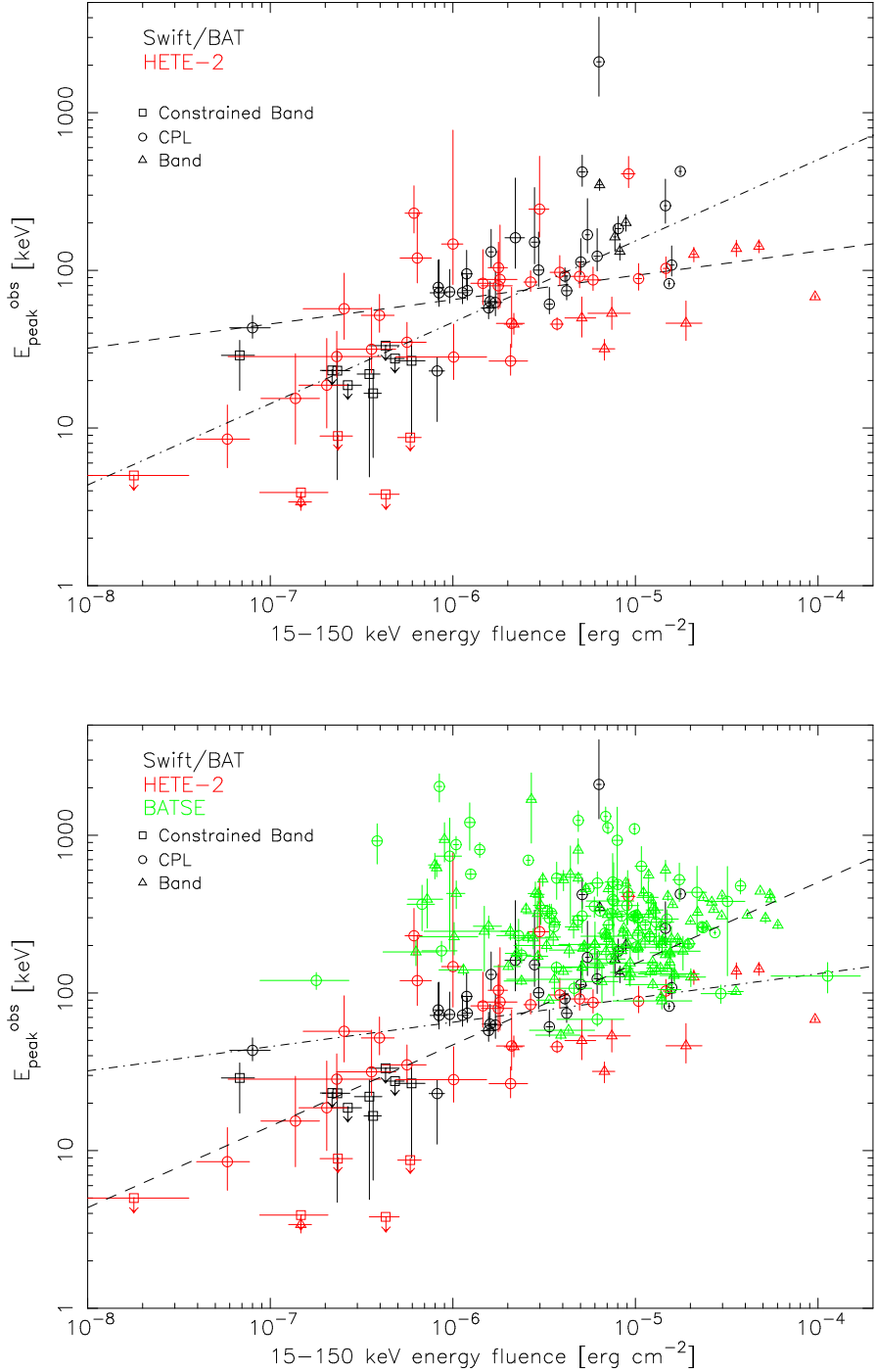

\centerline{
\includegraphics[angle=-90,scale=.5]{f8a.eps}}
\vspace{1cm}
\centerline{
\includegraphics[angle=-90,scale=.5]{f8b.eps}}
\caption{Top: A plot of the 15 -- 150 keV fluence and peak spectral energy
 $\eop$ for BAT (black) and {\it HETE-2} (red) samples. Bottom: A plot of 
the 15 -- 150 keV fluence and peak spectral energy
 $\eop$ for BAT (black), {\it HETE-2} (red) and BATSE (green) samples. 
The dashed and dash-dotted line are the best fit to the BAT and the {\it HETE-2} 
data with and without taking into account the errors, and are given by 
$\log(\eop) = 2.74_{-0.08}^{+0.15} + (0.15 \pm 0.02) \log(S(15-150~{\rm keV}))$ and 
$\log(\eop) = (4.77 \pm 0.63) + (0.52 \pm 0.11) \log(S(15-150~{\rm keV}))$.  
\label{ep_stot_all}}
\end{figure}

Figure~\ref{ep_stot_all} shows $\eop$ and S(15--150 keV) 
of the BAT, the {\it HETE-2} and the BATSE samples.  
The fluence in the 15--150 keV band for the {\it HETE-2} and 
the BATSE samples is calculated using the best fit spectral model reported 
in the catalog \citep[][]{sakamoto,kaneko2006}.  
The error in the fluence for the {\it HETE-2} and the BATSE samples is 
calculated by scaling the error in the normalization of the 
best fit spectral model.  
As clearly seen in the figure, S(15--150 keV) and $\eop$ of the BAT GRBs 
are consistent with both the {\it HETE-2} and the BATSE samples.  The strong 
correlation between $\eop$ and S(15--150 keV) still exists by combining 
the BAT and the {\it HETE-2} samples.  The correlation 
coefficient combining the BAT and the {\it HETE-2} GRBs is +0.685 for 83 samples.  
The probability of such a correlation occurring by chance is $<10^{-11}$.  
The best fit correlation function 
between $\eop$ and S(15--150 keV) with and without taking into account the errors are 
$\log(\eop) = 2.74_{-0.08}^{+1.51} + (0.15 \pm 0.02) \log[S(15-150~{\rm keV})]$ 
and
$\log(\eop) = (4.77 \pm 0.63) + (0.52 \pm 0.11) \log[S(15-150~{\rm keV})]$, respectively.  
However, 
as clearly shown in both Figure \ref{ep_hist} and \ref{ep_stot_all}, 
the BAT XRFs are not softer (or weaker) than the {\it HETE-2} sample.  This is 
because of the higher observed energy band of the BAT compared to that of the 
{\it HETE-2} Wide-field X-ray Monitor (WXM; 2--25 keV) \citep{shirasaki2003}.  
Thus, caution might be needed for comparing the BAT and the {\it HETE-2} 
XRF samples.
It is also clear from the figures that the $\eop$ distribution of the BATSE sample 
is systematically higher compared with the GRB samples of the {\it HETE-2} and the BAT 
because of lacking sensitivity below 20 keV for BATSE.  

\begin{figure}
\includegraphics[angle=-90,scale=.65]{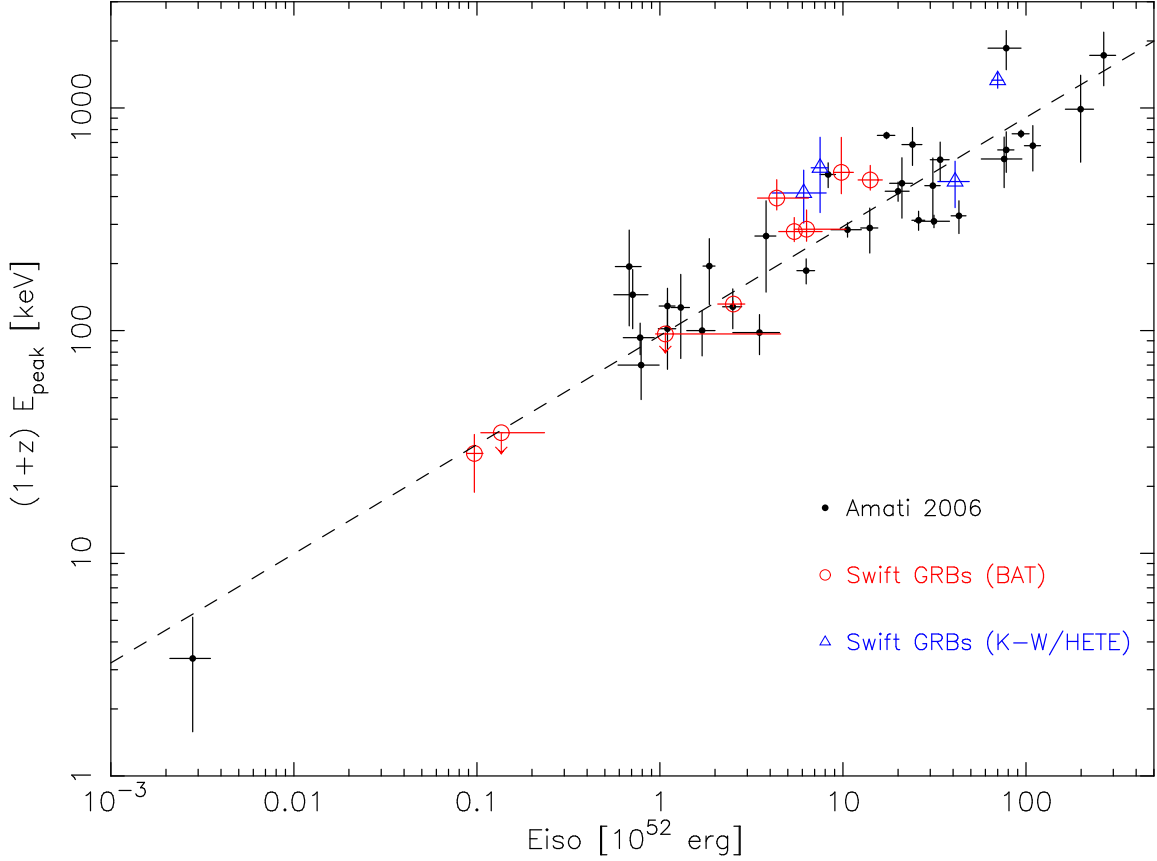}
\caption{Isotropic equivalent energy, $\eiso$ vs. the peak energy in the 
GRB rest frame, $\esp$ for the known redshift BAT GRBs in this work 
(red circles), 
pre-{\it Swift} GRBs (black dots) and the known redshift 
{\it Swift} GRBs observed by $Konus$-$Wind$ or {\it HETE-2} (blue triangles).  
The dashed line is the best fit correlation reported by 
 \citet{amati2006} ($\esp = 95~{\rm keV} \times \left(\frac{\eiso}{10^{52}~{\rm ergs}}\right)^{0.49}$).  
\label{ep_eiso}}
\end{figure}

Figure~\ref{ep_eiso} shows the correlation between the peak energy in the GRB 
rest frame $\esp$ ($\equiv$ (1+z) $\eop$) and the isotropic radiated 
energy $\eiso$.  We calculated $\esp$ and $\eiso$ for the 
nine known redshift GRBs\footnote{We exclude GRB 060512 
because of a less secure measurement of its redshift.} in our sample using the 
BAT data (Table \ref{tab:ep_eiso}).  For these GRBs, $\eiso$ is derived 
directly from the spectral fitting using the Band function and integrating 
from 1 keV to 10 MeV at the GRB rest frame.  $\esp$ is calculated from 
$\eop$ based on a CPL fit.  $\esp$ and $\eiso$ values for the remaining {\it Swift} 
GRBs are extracted from \citet{amati2006}.   The values for the pre-{\it Swift} 
GRBs are also extracted from \citet{amati2006}.  Although our sample of 
known redshift GRBs is small, we have confirmed the existence and 
the extension of the $\esp-\eiso$ relation to XRFs and XRRs 
(GRBs with $\esp < 100$ keV) for the {\it Swift} GRBs 
\citep{amati2002,lamb2005,sakamoto2,sakamoto2006}.  

\section{X-ray Afterglow Emission}

The spectral and temporal properties of the 41 X-ray afterglows are 
summarized in Tables \ref{tab:xrt_spec} and \ref{tbl:xray_temp}.  

\begin{figure}
\includegraphics[angle=-90,scale=.65]{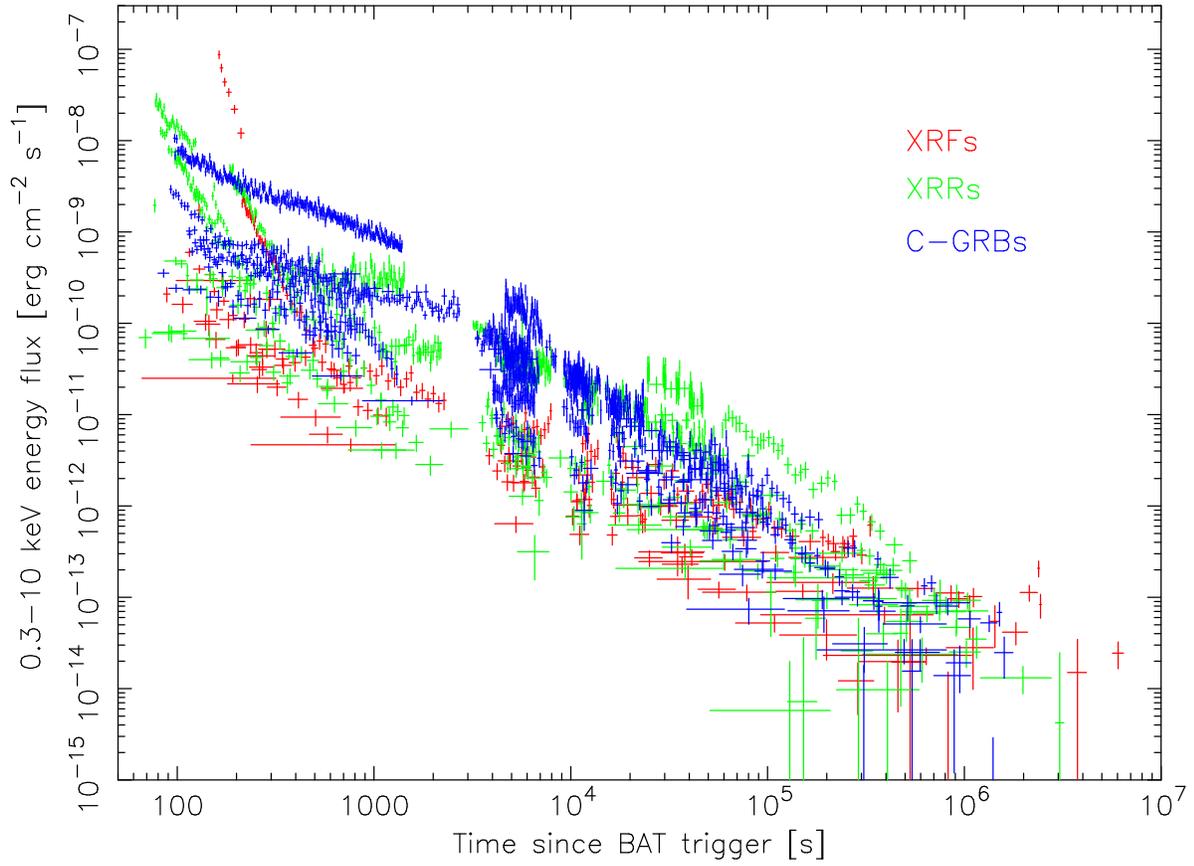}
\caption{A composite plot of the 0.3 -- 10 keV fluxes of the
X-ray afterglow light curves of the XRFs (red), XRRs (green), 
and C-GRBs (blue) in our sample. \label{xrt_lc_all}}
\end{figure}

\begin{figure}
\includegraphics[angle=-90,scale=.65]{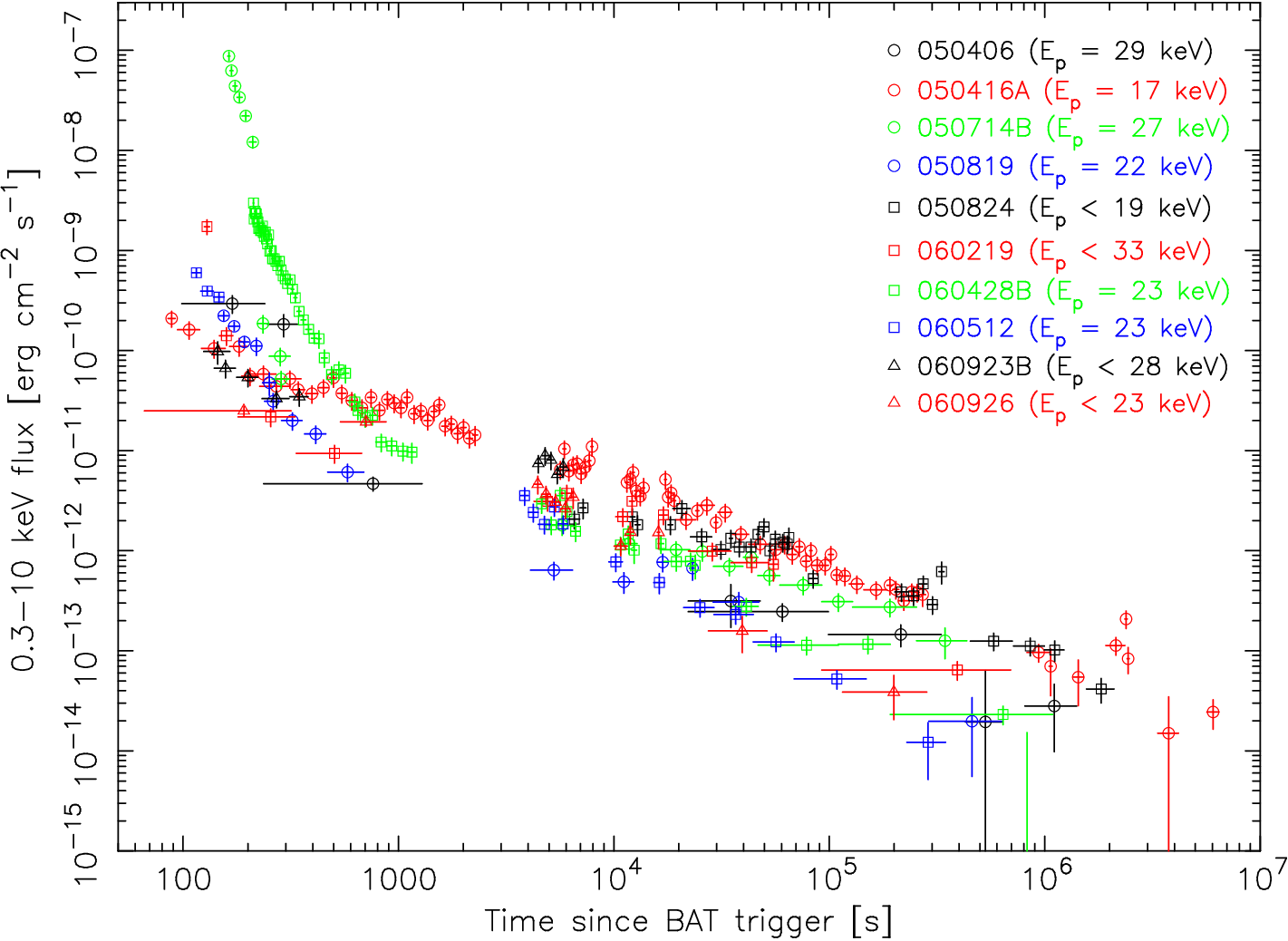}
\caption{A composite plot of the 0.3 -- 10 keV 
X-ray afterglow light curves of XRFs.  \label{xrt_lc_xrf}}
\end{figure}

\begin{figure}
\includegraphics[angle=-90,scale=.65]{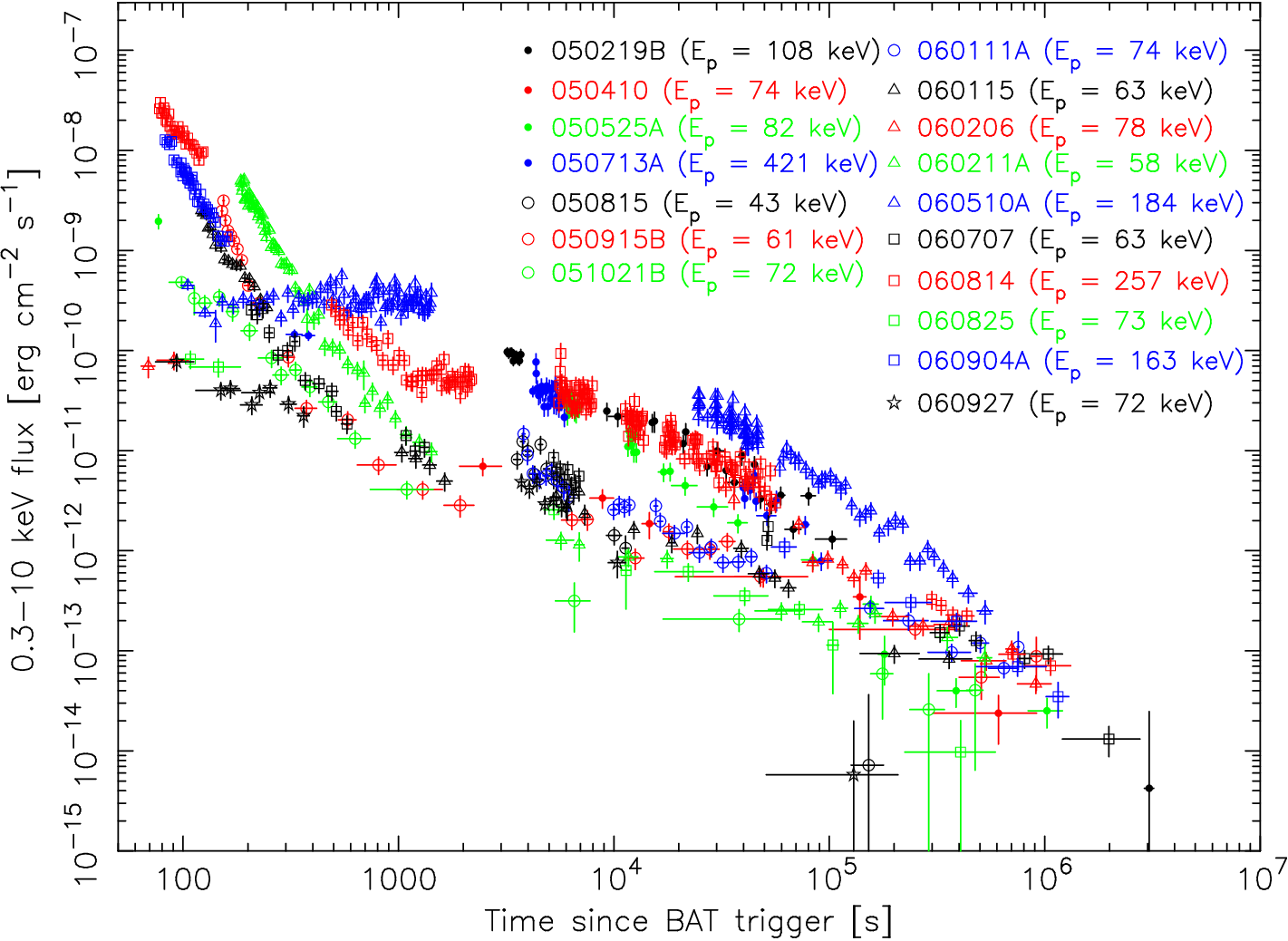}
\caption{A composite plot of the 0.3 -- 10 keV 
X-ray afterglow light curves of XRRs.   \label{xrt_lc_xrr}}
\end{figure}

\begin{figure}
\includegraphics[angle=-90,scale=.65]{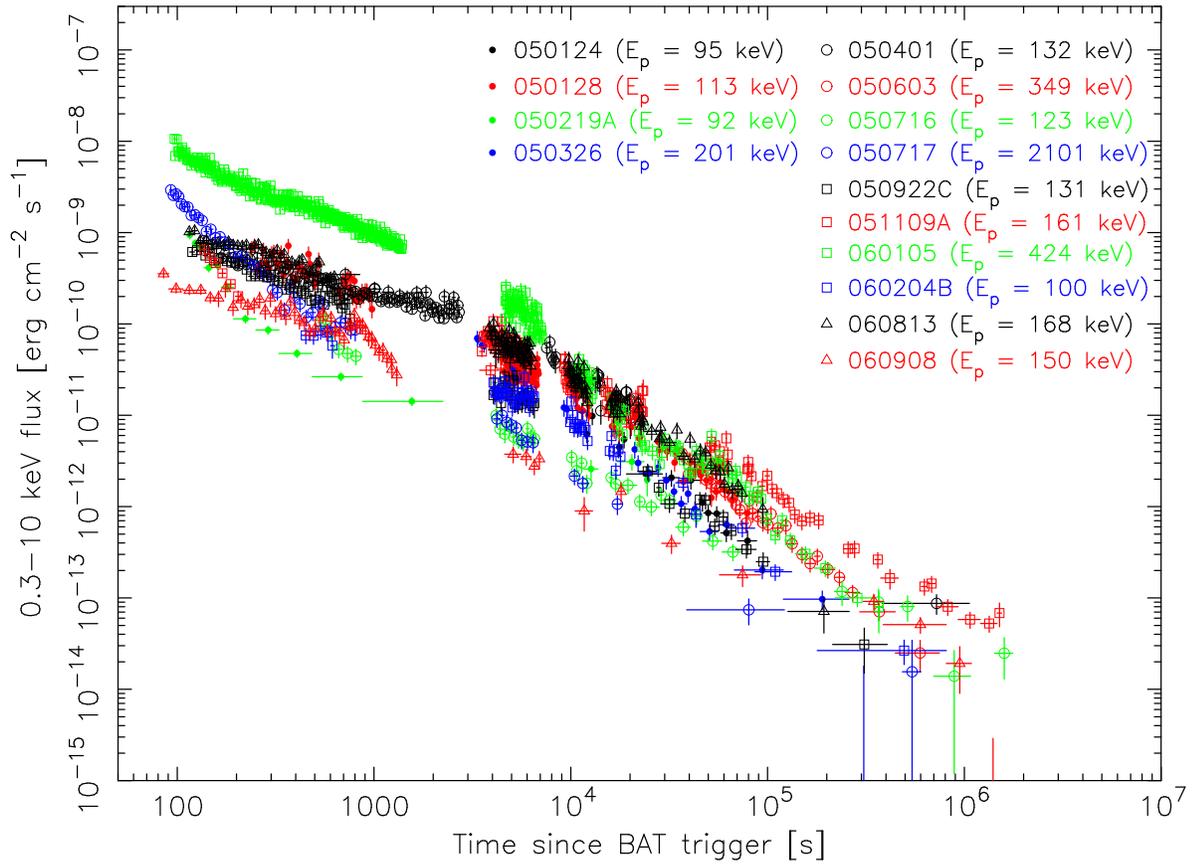}
\caption{A composite plot of the 0.3 -- 10 keV 
X-ray afterglow light curves of C-GRBs.   \label{xrt_lc_grb}}
\end{figure}

Figure~\ref{xrt_lc_all} is a composite plot of the X-ray afterglow light curves.  
Figures~\ref{xrt_lc_xrf},~\ref{xrt_lc_xrr}, and~\ref{xrt_lc_grb} show the light
curves in each GRB class.  As we subsequently discuss in detail, we find 
that C-GRBs in our sample tend to have afterglows with shallow 
decay indices at early times followed by steeper indices 
at later times, and that the breaks between these two indices occur at 
about $10^{3} - 10^{4}$ seconds.  On the other hand, XRF afterglows show a fairly shallow 
decay index until the end of the XRT observation 
without any significant break.  XRRs in our sample were split between 
these two behaviors, with some manifesting a pattern like the XRF sample 
and others a pattern like the C-GRB sample. 

\begin{figure}
\includegraphics[angle=-90,scale=.65]{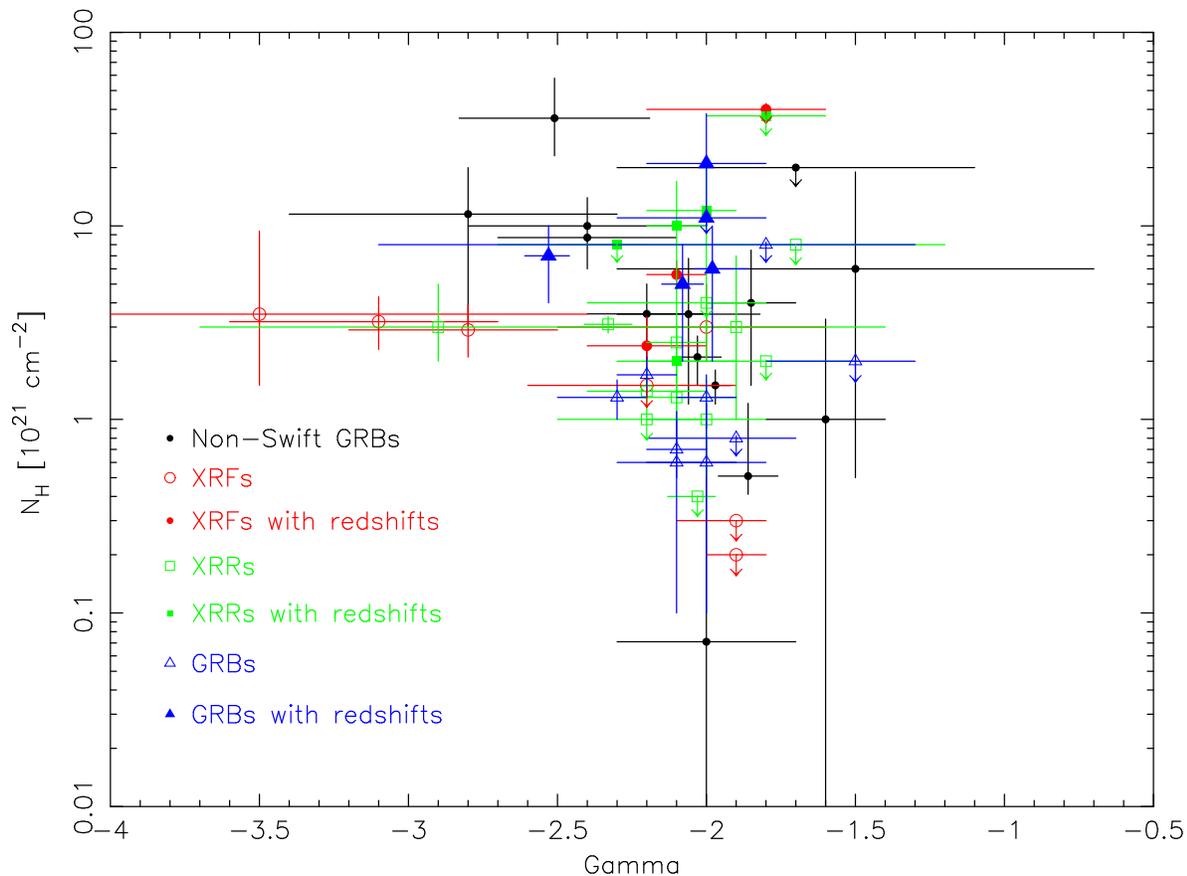}
\caption{A plot of the best-fit neutral hydrogen column densities $N_H$ and
photon indices $\Gamma$ of X-ray afterglows in our sample, along with values 
taken from \citet{frontera}.  The values plotted here of the 
{\it Swift} sample are taken from the 
PC mode spectra.  {\it Swift} XRFs, XRRs, C-GRBs and non-{\it Swift} samples are 
shown in red circles, green squares, blue triangles and black dots,  
respectively.  
\label{nh_gamma}}
\end{figure}

Figure~\ref{nh_gamma} shows the distribution of best-fit excess neutral 
hydrogen column densities $N_H$ over the galactic $N_H$ \citep{dickey1990} 
and photon indices $\Gamma$ in the PC mode for our sample of bursts.  For known 
redshift GRBs, the excess $N_H$ is calculated in the GRB rest frame.  
Also shown are the Beppo{\it SAX} values gathered and cited by \citet{frontera} 
for comparison.  There is no systematic differences in $N_H$ and $\Gamma$ between 
either the BAT and the pre-{\it Swift} GRBs or between the individual
classes of the BAT 
GRBs.  We also confirmed a significant amount of an excess $N_H$ for most of 
our sample \citep[e.g.][]{campana2006,grupe2007}. 

\begin{figure}
\includegraphics[angle=-90,scale=.65]{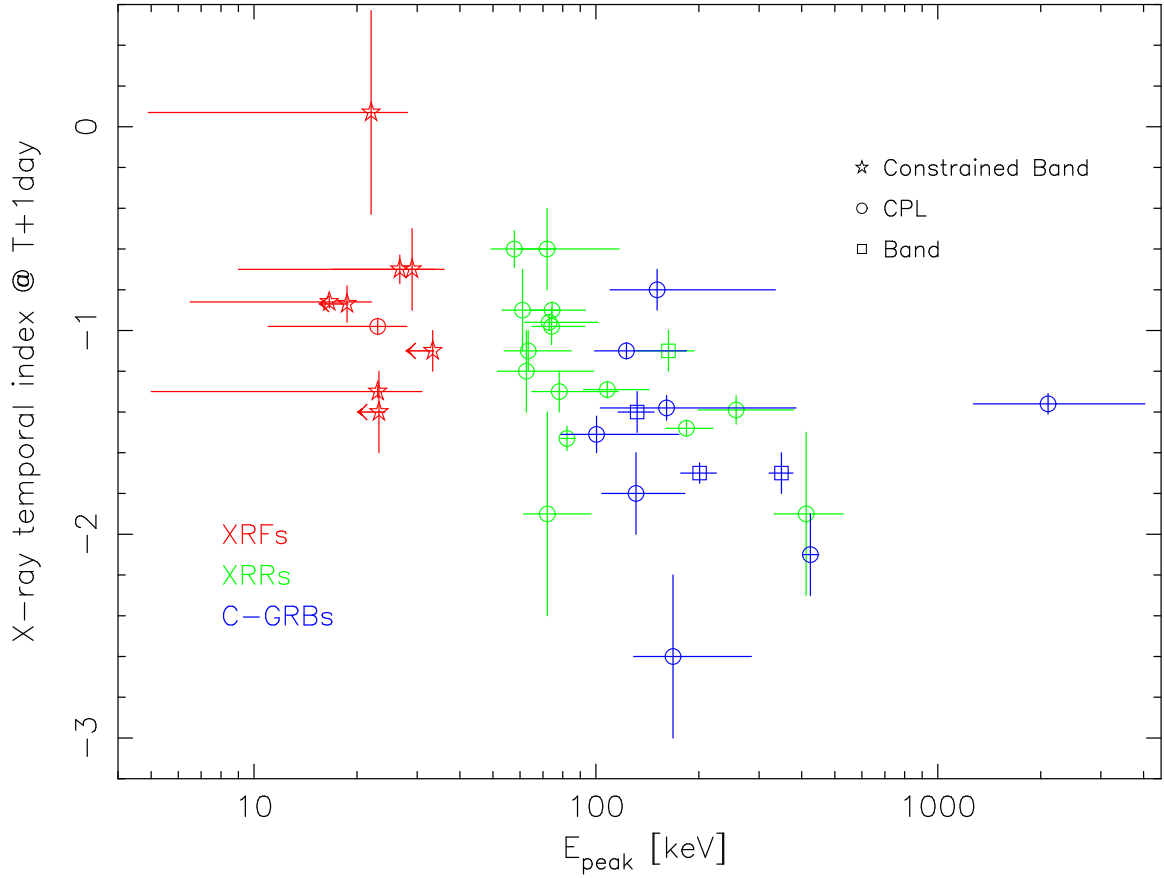}
\caption{A plot of the temporal decay indices measured 1 day
after the burst and $\eop$ of XRFs (red), XRRs (green) and C-GRBs (blue).  
$\eop$ values derived from 
a $constrained$ Band function, a CPL, and the Band function are marked 
as stars, circles, and squares, respectively. \label{eop_1d_decay}}
\end{figure}

Figure ~\ref{eop_1d_decay} shows the X-ray temporal index  
in the 0.3--10 keV band taken 1 day after the burst ($\alpha_{1 \rm day}$) 
plotted against $\eop$ for 36 bursts\footnote{We exclude GRB 050124, 
GRB 050128, GRB 050219A, GRB 050815, and GRB 060923B for this study because 
there are no X-ray data around 1 day after the burst.}.  
There is a systematic trend in $\alpha_{1 \rm day}$ of XRFs, in that they 
are concentrated around $-1$ and only one sample is steeper than $-1.5$.  
On the other hand, $\alpha_{1 \rm day}$ of XRRs and C-GRBs 
are much more widely spread.  Moreover, there might be a hint that XRRs 
and C-GRBs have a systematically steeper $\alpha_{1 \rm day}$ than XRFs.  
The correlation coefficient between $\alpha_{1 \rm day}$ and $\eop$ 
has been calculated using the same method for which we apply to calculate 
the correlation coefficient between $\eop$ and the fluence in the 1--1000 
keV band (section 3).  We investigate the correlation coefficient for 
1) GRBs without XRFs and GRB 050717 which is outlier with $\eop$ of 2 MeV 
(sample A; total 26 GRBs), 2) GRBs without GRB 050717 (sample B; total 35 GRBs), 
and 3) all 36 GRBs (sample C) to evaluate the systematic effect due to
significantly low or high $\eop$ values compared with the rest of the
samples.  We find the correlation coefficients of 
$-0.44_{-0.08}^{+0.07}$, $-0.44_{-0.07}^{+0.04}$, and $-0.49_{-0.06}^{+0.04}$ 
(all 1$\sigma$ errors) for samples A, B, and C respectively.  The probabilities 
of a chance occurrence in each sample size are $7.1 \times 10^{-3} - 6.7 \times 
10^{-2}$, $2.0 \times 10^{-3} - 1.9 \times 10^{-2}$, and $5.9 \times 10^{-4} 
- 6.3 \times 10^{-3}$ in the 1$\sigma$ interval for samples A, B, and C respectively.  
Therefore, if we include the XRF sample, the correlation between 
$\alpha_{1 \rm day}$ and $\eop$ is significant at the $>$99.98 \% level.  

\begin{figure}
\includegraphics[angle=-90,scale=.65]{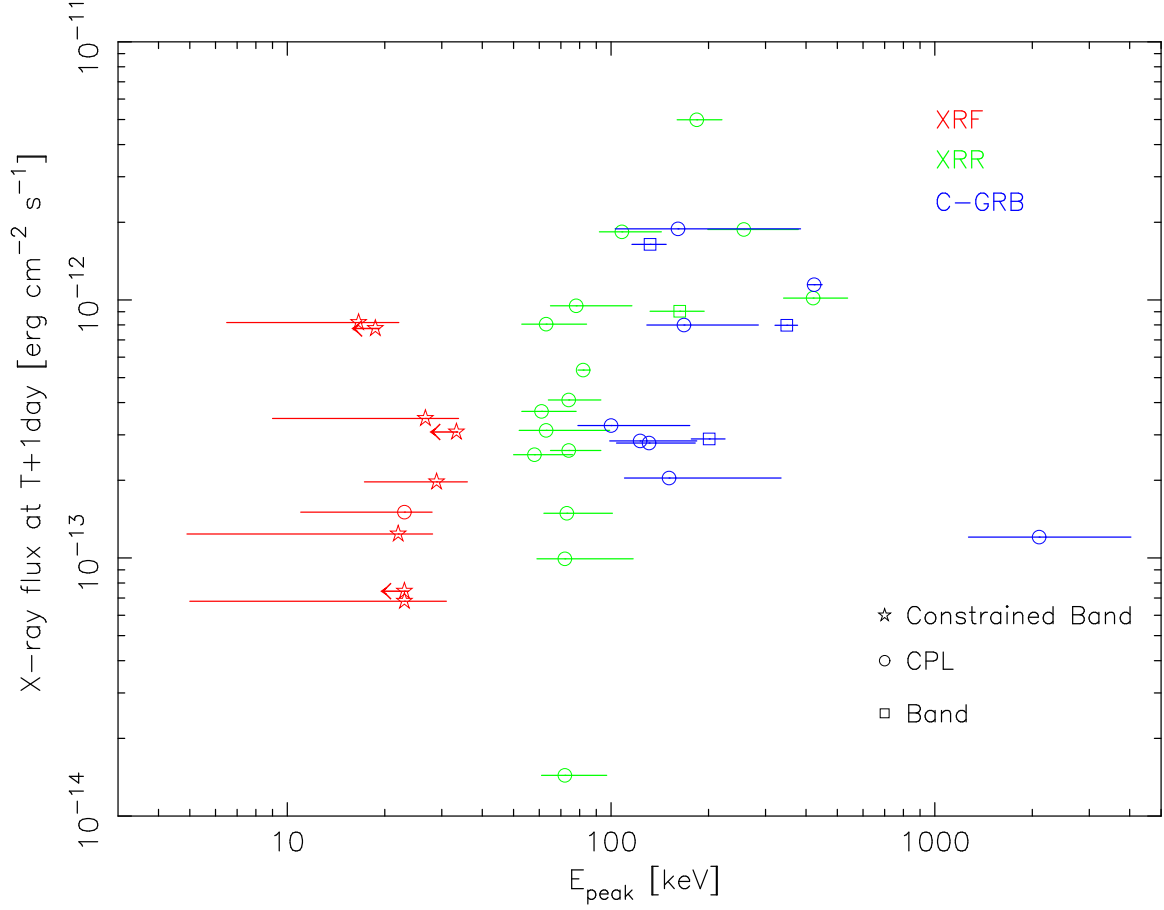}
\caption{A plot of the X-ray unabsorbed flux measured 1 day after the 
burst and $\eop$ of XRFs (red), XRRs (green) and C-GRBs (blue).  
$\eop$ values derived from a $constrained$ Band function, a CPL, and 
the Band function are marked as stars, circles, and squares, 
respectively. \label{eop_ag_flux_1d}}
\end{figure}

The relationship between the unabsorbed X-ray afterglow flux at 1 day after 
the burst and $\eop$ is shown in Figure \ref{eop_ag_flux_1d}.  We 
calculate the correlation coefficient between the X-ray flux and $\eop$ 
by the same method and also for the same three samples as we used in the
correlation study between $\alpha_{1 \rm day}$ 
and $\eop$ (Figure ~\ref{eop_1d_decay}).  The calculated correlation coefficients 
are $+0.48_{-0.07}^{+0.03}$, $+0.44_{-0.04}^{+0.05}$, and $+0.31_{-0.04}^{+0.04}$ 
(all 1$\sigma$ errors) for samples A, B, and C respectively.  The chance probabilities 
are $3.5 \times 10^{-2} - 8.5 \times 10^{-3}$, $1.9 \times 10^{-2} - 3.2 
\times 10^{-3}$ and $1.2 \times 10^{-1} - 3.9 \times 10^{-2}$ in 1$\sigma$ interval 
for samples A, B, and C respectively.  Therefore, there is no 
significant correlation between the X-ray flux and $\eop$ if we investigate 
for the whole 36 bursts (sample C).  However, the correlation becomes significant if we 
exclude GRB 050717, which is an outlier with $\eop$ of 2 MeV.  
Therefore, there might be a hint of a correlation between the X-ray 
flux at 1 day after the burst and $\eop$.  

\begin{figure}
\includegraphics[angle=-90,scale=.65]{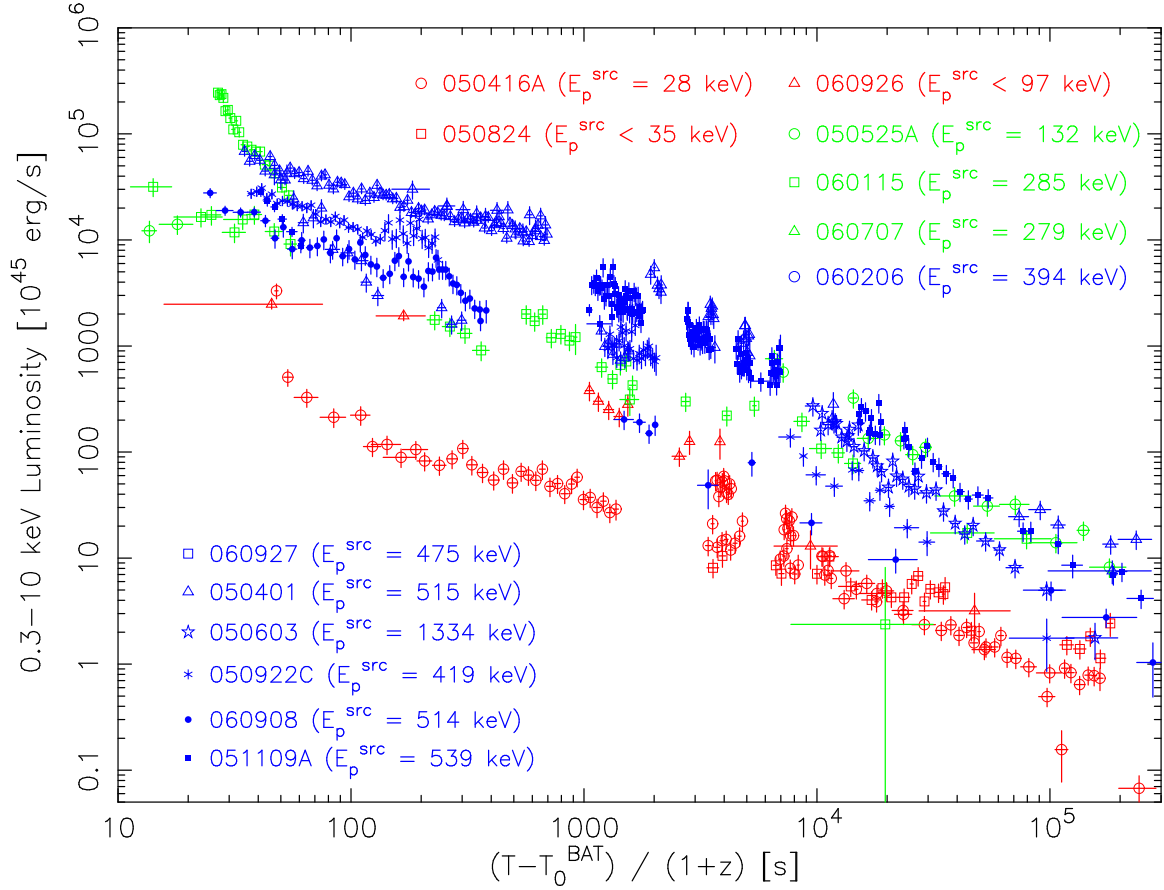}
\caption{The composite X-ray luminosity afterglow light curves for known 
redshift GRBs in our sample.  GRBs with $\esp < 100$ keV, 
$100$ keV $<$ $\esp < 300$ keV, and $\esp > 300$ keV are shown 
in red, green, and blue, respectively.  T$_{0}^{\rm BAT}$ refers to 
the BAT trigger time.  \label{xrt_lc_src}}
\end{figure}

\begin{figure}
\includegraphics[angle=-90,scale=.65]{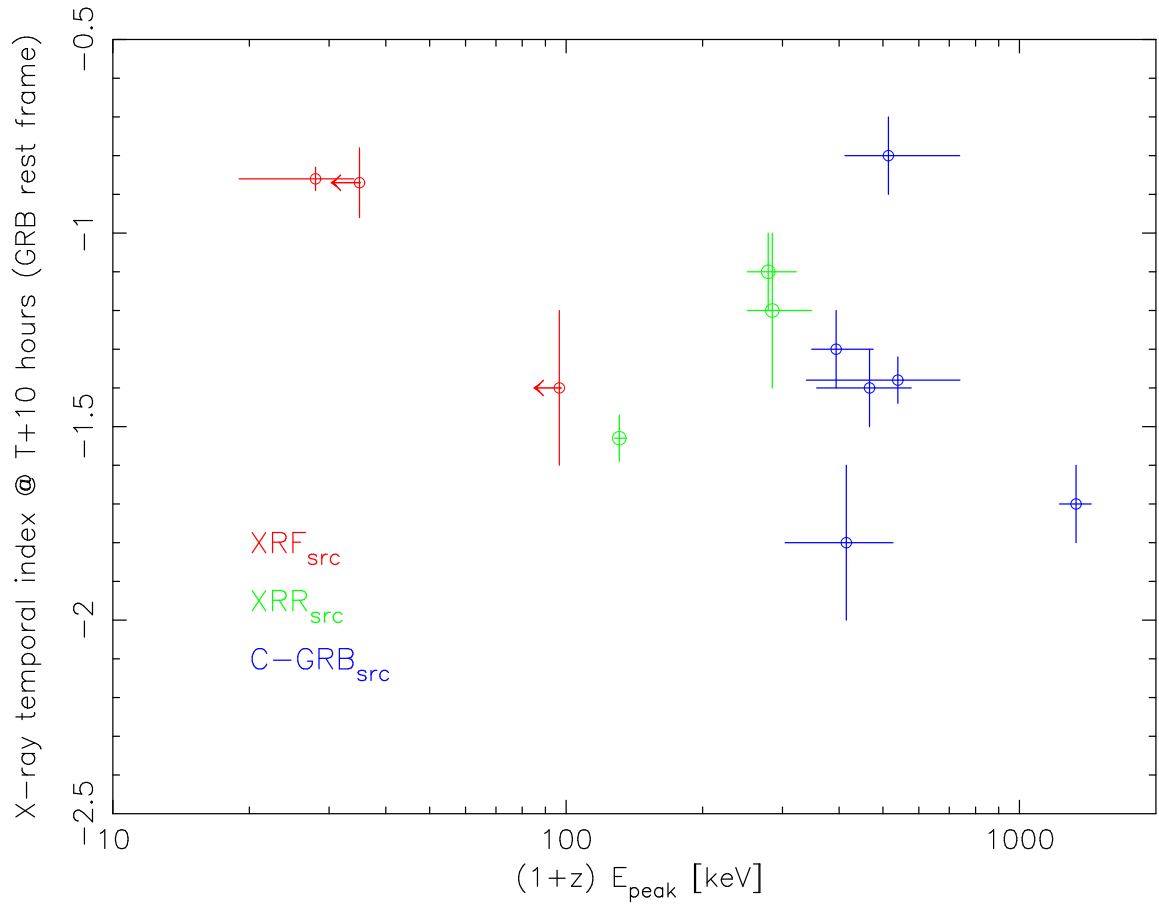}
\caption{A plot of the X-ray temporal index measured at 10 hours after the 
burst in the GRB rest frame and $\esp$.  \label{esp_10h_decay}}
\end{figure}

\begin{figure}
\includegraphics[angle=-90,scale=.65]{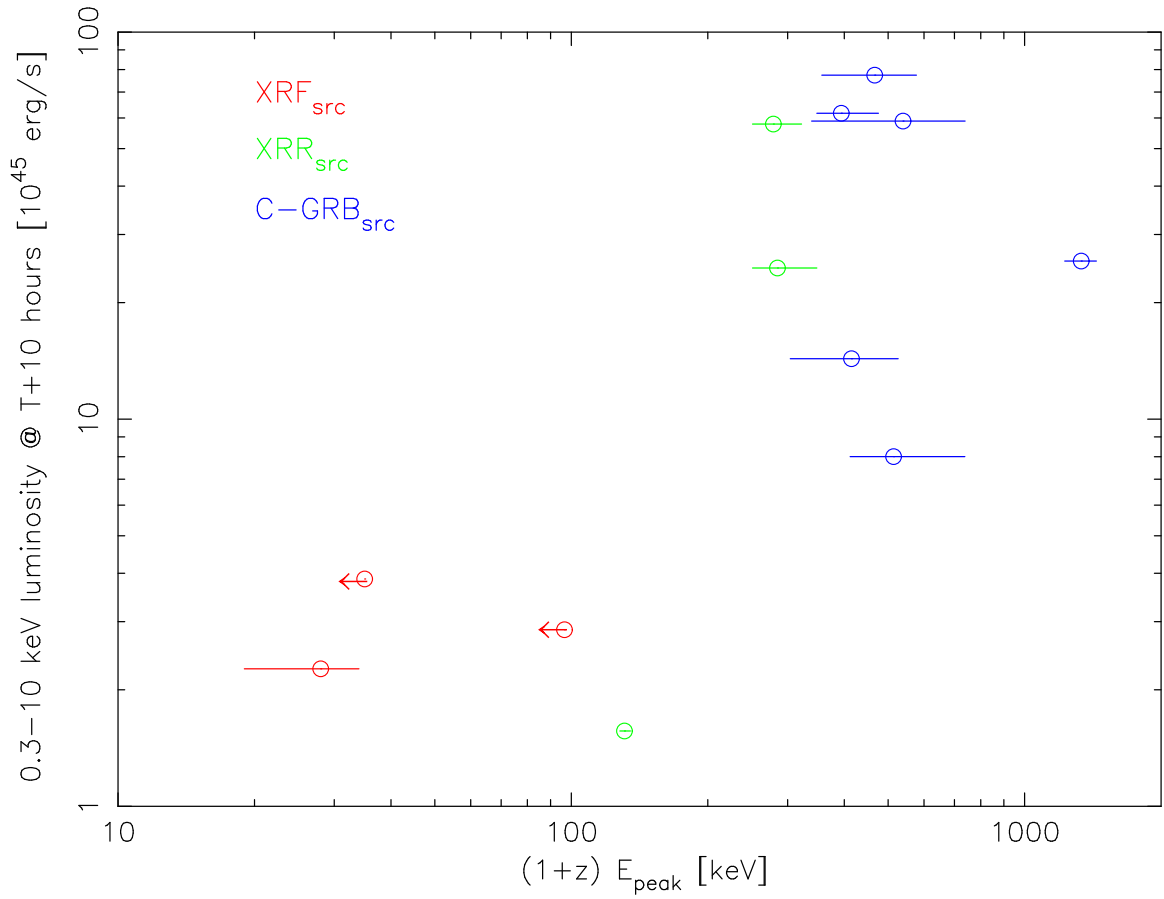}
\caption{A plot of the 0.3--10 keV luminosity measured at 10 hours after the 
burst in the GRB rest frame and $\esp$.
\label{esp_10h_lumi}}
\end{figure}

Figure~\ref{xrt_lc_src} shows the composite X-ray luminosity 
light curves for the known 
redshift GRBs in our sample.  The k-correction\footnote{The 0.3--10 keV 
luminosity, L$_{0.3-10}$, is calculated by 
L$_{0.3-10}$ $ = 4 \pi d_{L}^{2} (1+z)^{-\Gamma - 2} F_{0.3-10}$, 
where $d_{L}$ is the luminosity distance, $\Gamma$ is the photon index 
of the XRT spectra (Table \ref{tab:xrt_spec}) and $F_{0.3-10}$ is the 
observed flux in the 0.3-10 keV band.} has been applied to derive the 
0.3--10 keV luminosities from the X-ray fluxes of each light curve bin 
using the best fit PL photon index of the WT and the PC mode spectra.  The 
time dilation effect of the cosmic expansion is taken into account 
in these light curves.  The colors in the light curves are coded in the 
following ways: $\esp$ $<$ 100 keV in red (hereafter, XRF$_{\rm src}$, 
as XRF in the GRB rest frame), 100 keV $<$ $\esp$ $<$ 300 keV in green 
(hereafter, XRR$_{\rm src}$, as XRR in the GRB rest frame), and 
$\esp$ $>$ 300 keV in blue (hereafter, C-GRB$_{\rm src}$, as C-GRB in 
the GRB rest frame).  As illustrated in the figure, there are clear 
separations between XRF$_{\rm src}$, XRR$_{\rm src}$ and 
C-GRB$_{\rm src}$ in the overall luminosities of the X-ray light curves.  
XRFs$_{\rm src}$ have less luminosity by a factor of two or more 
compared to XRRs$_{\rm src}$ and C-GRBs$_{\rm src}$.  
Figure \ref{esp_10h_decay} and \ref{esp_10h_lumi} show the X-ray temporal 
index and the luminosity respectively at 10 hours after the burst in 
the GRB rest frame as a function of $\esp$.  
As seen in the observer's frame (Figure \ref{eop_1d_decay} and \ref{eop_ag_flux_1d}), 
there are weak correlations between $\esp$ and the temporal index and 
the luminosity.  
The correlation coefficients between $\esp$ and the temporal index, 
and between $\esp$ and the luminosity at 10 hours are $-0.53$ and 
$+0.72$ in both samples of 12\footnote{We exclude GRB 060927 because
there is no X-ray data around 10 hours at the GRB rest frame.}.  
The chance probabilities are 0.075 and 0.008.  
The global trend in the X-ray luminosity light curve 
is that XRFs$_{\rm src}$ have a temporal index of 
$\alpha \sim -1$ and smaller luminosities at 10 hours after the burst 
compared to those of XRRs$_{\rm src}$ and C-GRBs$_{\rm src}$.  

\section{Discussion}
\subsection{Characteristics between the prompt emission and the X-ray afterglow}

The results of our analysis strengthen the 
case that XRFs and long-duration
C-GRBs are not separate and distinct 
phenomena, but instead are simply ranges along a single continuum describing
some sort of broader phenomenon.
As Figure~\ref{sr_ep} illustrates, 
XRFs, XRRs and C-GRBs form a continuum
in peak energies $\eop$,
with XRF $\eop$ values tending to be
lower than those of XRRs,
which in turn are lower than those of C-GRBs.
Further evidence of the continuous nature of these phenomena
comes from the continuity 
in the fluences of XRFs, XRRs, and C-GRBs, 
with XRFs tending to manifest lower
fluences than XRRs, which tend to have lower fluences than C-GRBs.  This
is illustrated by the correlation between fluences and $\eop$ shown in
Figure~\ref{ep_stot}.  We also confirmed the existence of the extension 
of the $\esp$-$\eiso$ relation \citep{amati2002} to XRFs using our 
limited sample of known redshift GRBs.  

As we examine the X-ray afterglow properties of XRFs, XRRs and C-GRBs, 
we note that their spectral indices and natural hydrogen column densities 
show no strong correlation to indicate that the spectra of XRF afterglows 
are distinctly different from those of XRRs or C-GRBs.  We do, however, note a 
possible distinction in the shape of the afterglow light curves among XRFs, 
XRRs, and C-GRBs.  

\begin{figure}
\centerline{
\includegraphics[angle=0,scale=1]{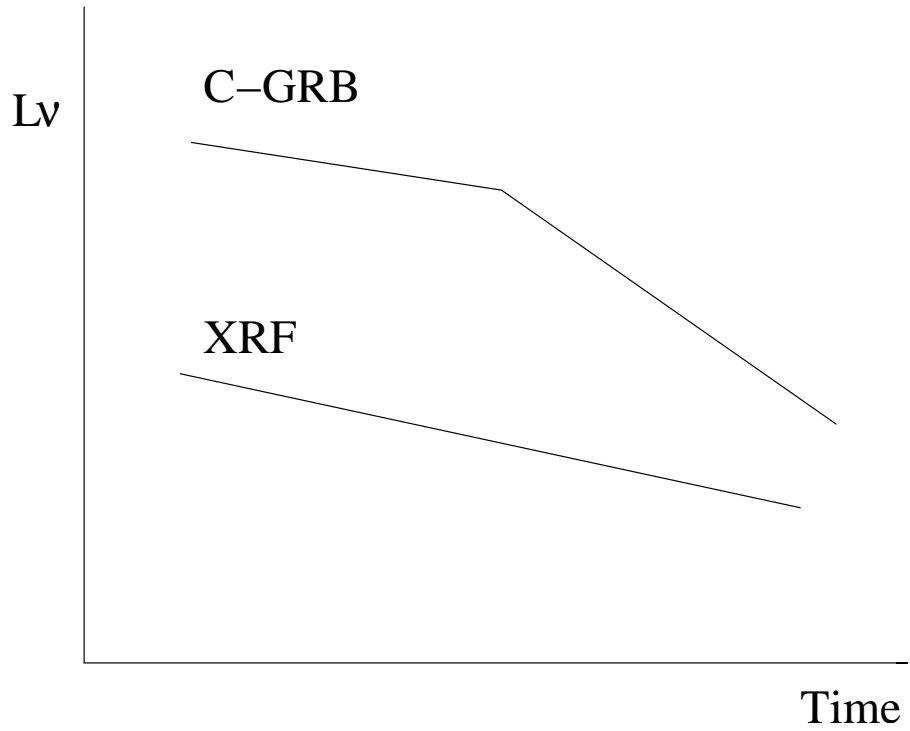}}
\caption{A schematic figure of XRF and C-GRB X-ray afterglow 
light curves.  C-GRB afterglows tend to have a shallow index followed by 
a steeper index, with a break around 10$^{3}-10^{4}$ seconds after the 
burst.  XRF afterglows, on the other hand, tend to have a shallow index 
without a break of significant change in the decay index.  Furthermore, 
the overall luminosity of XRF afterglows is factor of two or more less 
luminous than that of C-GRBs. \label{sche_fig_lc}}
\end{figure}

We find that the C-GRBs in our sample tend to have afterglows with shallow 
decay indices ($-1.3 < \alpha < -0.2$) at early times 
followed by steeper indices ($-2.0 < \alpha < -1.2$) 
at later times, and that the breaks between these two indices occur at 
about $10^{3} - 10^{4}$ seconds.   XRF afterglows, on the other hand, 
seem to follow a different pattern.  They often show a fairly shallow 
decay index ($-1.2 < \alpha <$ 0) until the end of the XRT observation 
without any significant break to $\alpha < -1.2$.  The afterglows of the 
XRRs in our sample were split between these two behaviors, 
with some manifesting a pattern like the XRF sample and others a pattern like 
the C-GRB sample (Figure \ref{xrt_lc_all}--\ref{xrt_lc_grb}).  It is possible that these two patterns form a continuum, 
with the break between shallow index and steep index occurring at later 
times for XRFs (sometimes after the afterglow has faded below our detection 
threshold) and at earlier times for C-GRBs (Figure \ref{eop_1d_decay}).   
There is, however, another possibility that 
this shallow-to-steep decay only exists in high $\ep$ GRBs.  
Furthermore, using our limited known redshift GRB sample, we confirmed our 
findings of the global features of the X-ray afterglows in the X-ray light 
curves in the 
GRB rest frame (Figure \ref{esp_10h_decay} and \ref{esp_10h_lumi}).  
Thus, the transition from a shallow to steep decay around 
$10^{3}-10^{4}$ seconds commonly seen in XRT light curves might somehow 
be related to the $E_{\rm peak}$ of its prompt emission (Figure \ref{sche_fig_lc}).  
Note that, however, two C-GRBs, GRB 050716 and GRB 060908, show a relatively shallow 
decay index without breaks up to $10^{6} - 10^{7}$ seconds after the trigger, 
and thus have the same afterglow behaviors as XRFs.  

\subsection{Difference in the X-ray afterglow luminosities}

As noted by \citet{gendre2007}, we also found differences in the 
luminosity of the X-ray light curves measured in the GRB rest frame.  
The luminosity of the global X-ray light curve is brighter when $\esp$ 
is higher (Figure \ref{xrt_lc_src}).  According to \citet{liang2006a}, 
there are two categories in the luminosity evolution of the optical 
afterglow.  They found that the dim group (having optical luminosities at 1 day 
of $\sim$ $5.3 \times 10^{44}$ ergs s$^{-1}$) all appear at redshifts lower 
than 1.1.  Motivated by their finding, we investigated $\esp$ of the \citet{liang2006a} 
sample using the values quoted in \citet{amati2006}.  We noticed 
that the $\esp$ values from their dim group are $<$ 200 keV.  The average 
$\esp$ of their dim group is 96 keV which would be XRFs$_{\rm, src}$ in our 
classification.  On the other hand, the average $\esp$ values from the bright 
group in their sample is 543 keV.  
Therefore, the trend which we found in the overall luminosity of 
the X-ray light curves might be consistent with the optical light curves.  
However, the break from a shallow-to-steep decay in the X-ray light curve 
which preferentially we see in C-GRBs is not usually observed 
in the optical band \citep[e.g.,][]{panaitescu2006}.  These similar 
and distinct characteristics in the X-ray and the optical afterglow 
light curves, together with the correlation in $\esp$, are important 
characteristics in seeking to understand the nature of the 
shallow-to-steep decay component in the X-ray afterglow data.  

\subsection{Understanding the shallow-to-steep decay by geometrical jet models}

There are several theoretical models which explain a shallow-to-steep 
decay break.  They are 1) the energy injection from the central engine 
or late time internal shocks 
\citep[e.g.,][]{nousek2006,zhang2006,ghisellini2007,panaitescu2007}, 
2) the geometrical jet models \citep[e.g.][]{eichler2006,toma2006}, 
3) the reverse shock \citep{genet2007,uhm2007}, 4) time-varying 
micro-physical parameters of the afterglow \citep{ioka2006}, or 
5) the dust scattering of prompt X-ray emission \citep{shao2007}.  
Here we focus on the geometrical jet models which 
have a tight connection between the prompt and afterglow 
emission properties.  \citet{eichler2006} 
investigated a thick ring jet (cross section of a jet in the shape 
of a ring) observed at slightly off-axis from the jet.  They 
can reproduce the shallow-to-steep decay feature in the X-ray afterglow 
with their thick ring jet model with the appearance of an 
off-axis afterglow emission at late times.  
Because of the relativistic beaming effect in this model, 
the observer, who is observing the ring jet from an off-axis direction, 
should see a softer prompt emission.  Therefore, 
we would expect to see a shallow-to-steep decay in the X-ray light
curve more frequently for XRFs and rarely for C-GRBs.  Our findings 
contradict this prediction of the model.  Another jet model which can 
produce a shallow-to-steep decay light curve is an 
inhomogeneous jet model \citep{toma2006}.  A shallow-to-steep decay 
phase of the light curve may be produced by the superposition of the sub-jet 
emissions which are launched slightly off-axis from the observer.  
The prediction of this jet model is that a shallow-to-steep decay
should co-exist with high $\esp$ in GRBs (an observer has to observe 
the prompt sub-jet emission from on-axis), and XRFs will have a 
conventional afterglow light curve.  
Our results agree quite nicely with this prediction.  
However, considering the non-existence of a 
shallow-to-steep phase in the optical light curve, it is hard to
understand why this shallow-to-steep phase only exists in the 
X-ray band in the framework of these jet models.  
Further simultaneous X-ray and optical afterglow observations 
along with a detailed modeling of afterglows taking into account the 
prompt emission properties such as $\ep$ will be needed to solve the 
origin of this mysterious shallow-to-steep decay feature.  

\section{Conclusion}

We have seen that the XRFs observed by {\it Swift} form a continuum with
the C-GRBs observed by {\it Swift} and by other missions, having systematically
lower fluences and lower $\eop$ than C-GRBs.  

We have noted that the X-ray light curves of XRFs tend to follow a different 
``template'' than those of C-GRBs.  The light curves of the C-GRB afterglows 
show a break to steeper indices (shallow-to-steep decay) at earlier 
times, whereas XRF afterglows show no such break.  
This break is evident in the X-ray but not 
in the optical light curve.  Moreover, the overall luminosity of XRF X-ray 
afterglows is smaller by a factor of two or more compared to that of C-GRBs.  
These distinct differences in the X-ray afterglow between XRFs and C-GRBs 
are keys to understanding not only the shallow-to-steep decay phase in the X-ray 
afterglow but also the nature of XRFs in an unified picture.  

We have discussed the geometrical jet models based on the trend 
which we found that the shallow-to-steep break in the X-ray afterglow 
preferentially is seen in the C-GRB sample.  
We concluded that none of the jet models can explain the behavior of a 
shallow-to-steep decay phase observed only in the X-ray afterglow.  We also emphasize the 
importance of having simultaneous X-ray and optical afterglow observations  
along with the characteristics of the prompt emission such as $\eop$ to 
constrain the various geometrical jet models.  

\acknowledgements

We would like to thank the anonymous referee for comments and suggestions
that materially improved the paper.  
This research was performed while T.S. participated in a NASA Postdoctoral 
Program
administered by Oak Ridge Associated Universities at NASA Goddard Space
Flight Center.  R.~Y. was supported in part by Grants-in-Aid for Scientific 
Research
of the Japanese Ministry of Education, Culture, Sports, Science,
and Technology 18740153.  
The material of the paper has been improved by
the discussions during the workshop ``Implications of {\it Swift's}
Discoveries about Gamma-Ray Bursts'' at the Aspen Center for Physics.

\clearpage



\input{tab1.tex}

\clearpage

\begin{deluxetable}{ccccc}
\tablecaption{$\esp$ and $\eiso$ derived from the BAT data.  
The uncertainty is 1$\sigma$. \label{tab:ep_eiso}}
\tablewidth{0pt}
\tablehead{
\colhead{GRB} &
\colhead{z} &
\colhead{$\esp$} &
\colhead{$\eiso$} &
\colhead{Instrument}\\
\colhead{} &
\colhead{} &
\colhead{(keV)} &
\colhead{($10^{52}$ erg)} &
\colhead{}}
\startdata
050401$^{1}$ & 2.9    & $467 \pm 110$ & $41 \pm 8$ & Konus-Wind\\ 
050416A & 0.6535 & $28_{-9}^{+6}$ & $0.096_{-0.009}^{+0.011}$ & BAT\\
050525A & 0.606  & $131_{-3}^{+4}$ & $2.5_{-0.5}^{+0.4}$ & BAT\\
050603$^{1}$ & 2.821  & $1333 \pm 107$ & $70 \pm 5$ & Konus-Wind\\
050824  & 0.83   & $<35$ & $0.13_{-0.03}^{+0.10}$ & BAT\\
050922C$^{1}$ & 2.198  & $415 \pm 111$ & $6.1 \pm 2.0$ & HETE-2\\
051109A$^{1}$ & 2.346  & $539 \pm 200$ & $7.5 \pm 0.8$ & Konus-Wind\\
060115  & 3.53   & $285_{-34}^{+63}$ & $6.3_{-0.9}^{+4.1}$ & BAT\\
060206  & 4.048  & $394_{-46}^{+82}$ & $4.3_{-0.9}^{+2.1}$ & BAT\\
060707  & 3.425  & $279_{-28}^{+43}$ & $5.4_{-1.0}^{+2.3}$ & BAT\\
060908$^{2}$ & 2.43 & $514_{-102}^{+224}$ & $9.8_{-0.9}^{+1.6}$ & BAT\\
060926  & 3.20   & $<96.6$ & $1.1_{-0.1}^{+3.5}$ & BAT\\
060927  & 5.6    & $475_{-47}^{+77}$ & $14.1_{-2.0}^{+2.3}$ & BAT\\
\enddata
\tablenotetext{1}{$\esp$ and $\eiso$ values from Amati (2006).}
\tablenotetext{2}{The high energy photon index $\Gamma_{2}$ of the Band 
function is fixed at $-2.3$.}
\end{deluxetable}

\input{tab3.tex}

\clearpage

\input{tab4.tex}

\clearpage

\clearpage

\clearpage

\clearpage

\clearpage

\clearpage

\clearpage

\clearpage

\clearpage

\clearpage

\clearpage

\clearpage

\clearpage

\clearpage

\clearpage

\clearpage

\clearpage

\clearpage

\clearpage

\clearpage

\end{document}

%% file: tab1.tex
\begin{deluxetable}{lc|ccc|cccc|cc|cc}
\tabletypesize{\scriptsize}
\rotate
\tablecaption{Prompt Emission Properties of 41 Swift Bursts\label{tbl1}}
\tablewidth{0pt}
\tablehead{
\colhead{GRB} & 
\colhead{T$^{\rm BAT}_{100}$$^{\ast}$} &
\multicolumn{3}{c}{PL} & 
\multicolumn{4}{c}{CPL} & 
\multicolumn{2}{c}{Other} & 
\colhead{S(15-150 keV)$^\P$} & 
\colhead{SR23$^{\Diamond}$}\\
\colhead{} & 
\colhead{} & 
\colhead{$\Gamma$} & 
\colhead{K$_{50}$$^{\star}$} & 
\colhead{$\chi^{2}$$^{\oplus}$} & 
\colhead{$\Gamma$} & 
\colhead{$\eop$$^{\diamond}$} & 
\colhead{K$_{50}$$^{\dagger}$} & 
\colhead{$\chi^{2}$$^{\oplus}$} & 
\colhead{$\eop$$^{\diamond}$} & 
\colhead{Mo/Inst$^{\ddagger}$} & 
\colhead{} & 
\colhead{}
}
\startdata
XRF 050406 & 6.4 & $-2.4_{-0.4}^{+0.3}$ & $13 \pm 4$ &  82.7 & -- & -- & -- & -- & $29_{-12}^{+7}$ & C-Band & $0.79 \pm 0.17$ & $1.3 \pm 0.6$\\
XRF 050416A & 3.0 & $-3.1 \pm 0.2$ & $98 \pm 17$ & 58.8 & -- & -- & -- & -- & $17_{-10}^{+6}$ & C-Band & $3.7 \pm 0.4$ & $2.1 \pm 0.6$\\
XRF 050714B & 50.3 & $-2.4 \pm 0.3$ & $12 \pm 3$ & 45.3 & -- &  -- & -- & -- & $27_{-18}^{+7}$ & C-Band & $6.0 \pm 1.1$ & $1.4 \pm 0.5$\\
XRF 050819 & 47.3 & $-2.7 \pm 0.3$ & $7 \pm 2$ & 57.1 & -- & -- & -- & -- & $22_{-17}^{+6}$ & C-Band & $3.5 \pm 0.5$ & $1.6 \pm 0.6$\\
XRF 050824 & 26.6 & $-2.8 \pm 0.4$ & $9 \pm 3$ & 49.9 & -- & -- & -- & -- & $<$19 & C-Band & $2.7 \pm 0.5$ & $1.6 \pm 0.8$\\
XRF 060219 & 65.3 & $-2.6_{-0.4}^{+0.3}$ & $6 \pm 2$ & 59.6 & -- & -- & -- & -- & $<$33 & C-Band & $4.3 \pm 0.8$ & $1.4 \pm 0.6$\\
XRF 060428B & 65.7 & $-2.6 \pm 0.2$ & $12 \pm 2$ & 66.7 & $-0.8_{-1.2}^{+1.7}$ & $23_{-12}^{+5}$ & $19_{-15}^{+210}$ & 59.1 & -- & -- & $8.2 \pm 0.8$ & $1.5 \pm 0.3$\\
XRF 060512 & 9.7 & $-2.5 \pm 0.3$ & $24 \pm 5$ & 36.1 & -- & -- & -- & -- & $23_{-18}^{+8}$ & C-Band & $2.3 \pm 0.4$ & $1.4 \pm 0.5$\\
XRF 060923B & 9.9 & $-2.5_{-0.3}^{+0.2}$ & $49 \pm 8$ & 56.9 & -- & -- & -- & -- & $<$27.6 & C-Band & $4.8 \pm 0.6$ & $1.4 \pm 0.4$\\
XRF 060926 & 8.7 & $-2.5 \pm 0.2$ & $25 \pm 4$ & 58.5 & -- & -- & -- & -- & $<$23 & C-Band & $2.2 \pm 0.3$ & $1.5 \pm 0.4$\\\hline
XRR 050219B & 76.1 & $-1.54 \pm 0.05$ & $224 \pm 7$ & 86.6 & $-1.0 \pm 0.2$ & $108_{-16}^{+35}$ & $39_{-8}^{+10}$ & 69.0 & -- & -- & $161 \pm 5$ & $0.73 \pm 0.04$\\
XRR 050410 & 49.5 & $-1.65 \pm 0.08$ & $94 \pm 4$ & 78.5 & $-0.8 \pm 0.4$ & $74_{-9}^{+19}$ & $24_{-8}^{+13}$ & 61.3 & -- & -- & $42 \pm 2$ & $0.78 \pm 0.06$ \\
XRR 050525A & 12.8 & $-1.76$ & 1350 & 166.4 & $-1.0 \pm 0.1$ & $82_{-3}^{+4}$ & $274_{-27}^{+30}$ & 17.9 & -- & -- & $153 \pm 2$ & 0.85\\
XRR 050713A & 190.7 & $-1.54 \pm 0.08$ & $28 \pm 1$ & 70.8 & -- & -- & -- & -- & $421_{-80}^{+117}$ & BAT/KW$^{1}$ & $51 \pm 2$ & $0.72 \pm 0.05$\\
XRR 050815 & 3.2 & $-1.8 \pm 0.2$ & $32 \pm 6$ & 75.6 & $0.9_{-1.4}^{+1.9}$ & $43_{-6}^{+9}$ & $130_{-111}^{+1390}$ & 62.1 & -- & -- & $0.8 \pm 0.1$ & $0.9 \pm 0.2$\\
XRR 050915B & 56.7 & $-1.90 \pm 0.06$ & $67 \pm 2$ & 55.5 & $-1.4 \pm 0.3$ & $61_{-8}^{+17}$ & $12_{-3}^{+4}$ & 46.0 & -- & -- & $33.8 \pm 1.4$ & $0.93 \pm 0.05$\\
XRR 051021B & 59.6 & $-1.55 \pm 0.14$ & $16 \pm 1$ & 56.9 & $-0.6_{-0.6}^{+0.8}$ & $72_{-13}^{+45}$ & $5_{-3}^{+7}$ & 49.7 & -- & -- & $8.4 \pm 0.9$ & $0.73 \pm 0.10$\\
XRR 060111A & 18.2 & $-1.65 \pm 0.07$ & $75 \pm 3$ & 69.0 & $-0.9 \pm 0.3$ & $74_{-10}^{+19}$ & $17_{-5}^{+8}$ & 50.4 & -- & -- & $12.0 \pm 0.6$ & $0.78 \pm 0.05$\\
XRR 060115 & 157.3 & $-1.7 \pm 0.1$ & $13 \pm 1$ & 52.6 & $-1.0_{-0.5}^{+0.6}$ & $63_{-11}^{+36}$ & $3_{-1}^{+3}$ & 45.8 & -- & -- & $17.1 \pm 1.5$ & $0.84 \pm 0.09$\\
XRR 060206 & 12.6 & $-1.71 \pm 0.08$ & $74 \pm 3$ &  64.6 & $-1.2 \pm 0.3$ & $78_{-13}^{+38}$ & $14_{-4}^{+6}$ & 55.3 & -- & -- & $8.3 \pm 0.4$ & $0.82 \pm 0.06$\\
XRR 060211A & 143.2 & $-1.8 \pm 0.1$ & $13 \pm 1$ & 71.5 & $-0.9_{-0.5}^{+0.6}$ & $58_{-8}^{+18}$ & $4_{-2}^{+4}$ & 60.6 & -- & -- & $15.7 \pm 1.4$ & $0.85 \pm 0.10$\\
XRR 060510A & 23.7 & $-1.57 \pm 0.07$ & $362 \pm 13$ & 54.0 & -- & -- & -- & -- & $184_{-24}^{+36}$ & KW$^{2}$ & $80.5 \pm 3.1$ & $0.74 \pm 0.05$\\
XRR 060707 & 74.4 & $-1.7 \pm 0.1$ & $25 \pm 2$ & 70.5 & $-0.6_{-0.6}^{+0.7}$ & $63_{-10}^{+21}$ & $9_{-4}^{+10}$ & 60.5 & -- & -- & $16.0 \pm 1.5$ & $0.8 \pm 0.1$\\
XRR 060814 & 230.3 & $-1.53 \pm 0.03$ & $67 \pm 1$ & 30.1 & -- & -- & -- & -- & $257_{-58}^{+122}$ & KW$^{3}$ & $146 \pm 2$ & $0.72 \pm 0.02$\\
XRR 060825 & 10.6 & $-1.72 \pm 0.07$ & $103 \pm 4$ & 64.8 & $-1.2 \pm 0.3$ & $73_{-11}^{+28}$ & $20_{-6}^{+9}$ & 53.7 & -- & -- & $9.6 \pm 0.5$ & $0.83 \pm 0.05$\\
XRR 060904A & 132.5 & $-1.55 \pm 0.04$ & $62 \pm 1$ & 43.6 & -- & -- & -- & -- & $163 \pm 31$ & KW$^{4}$ & $77.2 \pm 1.5$ & $0.73 \pm 0.02$\\
XRR 060927 & 24.7 & $-1.65 \pm 0.08$ & $52 \pm 2$ & 70.4 & $-0.9 \pm 0.4$ &  $72_{-11}^{+25}$ & $12_{-4}^{+7}$ & 57.5 & -- & -- & $11.3 \pm 0.7$ & $0.78 \pm 0.06$\\\hline
GRB 050124 & 6.0 & $-1.47 \pm 0.08$ & $213 \pm 11$ & 58.7 & $-0.7 \pm 0.4$ & $95_{-16}^{+39}$ & $47_{-15}^{+23}$ & 45.4 & -- & -- & $11.9 \pm 0.7$ & $0.69 \pm 0.06$\\
GRB 050128 & 30.4 & $-1.37 \pm 0.07$ & $172 \pm 7$ & 59.3 & $-0.7 \pm 0.3$ & $113_{-19}^{+46}$ & $35_{-10}^{+14}$ & 44.8 & -- & -- & $50.2 \pm 2.3$ & $0.65 \pm 0.05$\\
GRB 050219A & 35.1 & $-1.31 \pm 0.06$ & $123 \pm 4$ & 103.2 & $-0.1 \pm 0.3$ & $92_{-8}^{+12}$ & $41_{-10}^{+15}$ & 45.5 & -- & -- & $41.1 \pm 1.6$ & $0.62 \pm 0.03$\\
GRB 050326 & 41.0 & $-1.25 \pm 0.04$ & $216 \pm 4$ & 42.1 & -- & -- & -- & -- & $201 \pm 24$ & KW$^{5}$ & $88.6 \pm 1.6$ & $0.59 \pm 0.02$\\
GRB 050401 & 36.8 & $-1.40 \pm 0.07$ & $231 \pm 9$ & 37.1 & -- & -- & -- & -- & $132 \pm 16$ & KW$^{6}$ & $82.2 \pm 3.1$ & $0.66 \pm 0.04$\\
GRB 050603 & 21.4 & $-1.16 \pm 0.06$ & $289 \pm 10$ & 71.1 & -- & -- & -- & -- & $349 \pm 28$ & KW$^{7}$ & $63.6 \pm 2.3$ & $0.56 \pm 0.03$\\
GRB 050716 & 90.1 & $-1.37 \pm 0.06$ & $72 \pm 3$ & 52.5 & $-0.8 \pm 0.3$ & $123_{-24}^{+61}$ & $13_{-3}^{+4}$ & 39.4 & -- & -- & $61.7 \pm 2.4$ & $0.65 \pm 0.04$\\
GRB 050717 & 209.2 & $-1.30 \pm 0.05$ & $31 \pm 1$ & 48.5 & -- & -- & -- & -- & $2101_{-830}^{+1934}$ & BAT/KW$^{8}$ & $63.1 \pm 1.8$ & $0.61 \pm 0.03$\\
GRB 050922C & 6.8 & $-1.37 \pm 0.06$ & $247 \pm 8$ & 44.9 & -- & -- & -- & -- & $131_{-27}^{+51}$ & HETE$^{9}$ & $16.2 \pm 0.5$ & $0.65 \pm 0.03$\\
GRB 051109A & 45.4 & $-1.5 \pm 0.2$ & $51 \pm 6$ & 63.7 & -- & -- & -- & -- & $161_{-58}^{+224}$ & KW$^{10}$ & $22.0 \pm 2.7$ & $0.7 \pm 0.1$\\
GRB 060105 & 87.6 & $-1.07 \pm 0.04$ & $191 \pm 4$ & 32.5 & -- & -- & -- & -- & $424_{-20}^{+25}$ & KW$^{11}$ & $176 \pm 3$ & $0.53 \pm 0.02$\\
GRB 060204B & 195.0 & $-1.44 \pm 0.09$ & $17 \pm 1$ &  47.0 & $-0.8 \pm 0.4$ & $100_{-21}^{+75}$ & $3 \pm 2$ & 38.9 & -- & -- & $29.5 \pm 1.8$ & $0.68 \pm 0.05$\\
GRB 060813 & 36.7 & $-1.36 \pm 0.04$ & $155 \pm 3$ & 54.1 & $-1.0 \pm 0.2$ & $168_{-39}^{+117}$ & $22_{-4}^{+5}$ & 43.5 & -- & -- & $54.6 \pm 1.4$ & $0.64 \pm 0.03$\\
GRB 060908 & 28.5 & $-1.35 \pm 0.06$ & $103 \pm 3$ & 50.7 & $-1.0 \pm 0.3$ & $150_{-41}^{+184}$ & $15_{-3}^{+5}$ & 44.2 & -- & -- & $28.0 \pm 1.1$ & $0.63 \pm 0.04$\\
\enddata
\tablenotetext{\ast}{in seconds.}
\tablenotetext{\star}{in 10$^{-4}$ photons cm$^{-2}$ s$^{-1}$ keV$^{-1}$.}
\tablenotetext{\diamond}{in keV.}
\tablenotetext{\dagger}{in 10$^{-3}$ photons cm$^{-2}$ s$^{-1}$ keV$^{-1}$.}
\tablenotetext{\ddagger}{Spectral fitting model used/GRB instrument which reports $\eop$.}
\tablenotetext{\P}{BAT 15-150 keV energy fluence in 10$^{-7}$ ergs cm$^{-2}$ s$^{-1}$ with the BAT best fit model.}
\tablenotetext{\Diamond}{A fluence ratio of S(25-50 keV)/S(50-100 keV) derived from a PL fit.}
\tablenotetext{\oplus}{The degrees of freedom in a PL fit and a CPL fit are 57 and 56 respectively.}
\tablenotetext{1}{Morris et al. 2007;$\eop$ derived from a CPL model.}
\tablenotetext{2}{Golenetskii et al. 2006a, GCN Circ. 5113. $\eop$ derived from a CPL model.}
\tablenotetext{3}{Golenetskii et al. 2006c, GCN Circ. 5460. $\eop$ derived from a CPL model.}
\tablenotetext{4}{Golenetskii et al. 2006d, GCN Circ. 5518. $\eop$ derived from a Band model.}
\tablenotetext{5}{Golenetskii et al. 2005a, GCN Circ. 3152. $\eop$ derived from a Band model.}
\tablenotetext{6}{Golenetskii et al. 2005b, GCN Circ. 3179. $\eop$ derived from a Band model for the first episode.}
\tablenotetext{7}{Golenetskii et al. 2005c, GCN Circ. 3518. $\eop$ derived from a Band model.}
\tablenotetext{8}{Krimm et al. 2007;$\eop$ derived from a CPL model.}
\tablenotetext{9}{Crew et al. 2005, GCN Circ. 4021. $\eop$ derived from a CPL model.}
\tablenotetext{10}{Golenetskii et al. 2005d, GCN Circ. 4238. $\eop$ derived from a CPL model.}
\tablenotetext{11}{Tashiro et al. 2007;$\eop$ derived from a CPL model.}
\end{deluxetable}

%% file: tab3.tex
\begin{deluxetable}{lccccc|ccccc}
\tabletypesize{\scriptsize}
\rotate
\tablecaption{XRT X-ray spectral properties of 41 Swift Bursts\label{tab:xrt_spec}}
\tablewidth{0pt}
\tablehead{
\colhead{GRB} & 
\multicolumn{5}{c}{WT} & 
\multicolumn{5}{c}{PC}\\
\colhead{} & 
\colhead{t$_{\rm start}$} & 
\colhead{t$_{\rm stop}$} & 
\colhead{N$_{H}$} &
\colhead{$\Gamma^{\dagger}$} & 
\colhead{$\chi^{2}$/d.o.f.} & 
\colhead{t$_{\rm start}$} & 
\colhead{t$_{\rm stop}$} & 
\colhead{N$_{H}$} &
\colhead{$\Gamma^{\dagger}$} & 
\colhead{$\chi^{2}$/d.o.f.}\\
\colhead{} & 
\colhead{[s]} & 
\colhead{[s]} & 
\colhead{[10$^{21}$ cm$^{-2}$]} &
\colhead{} & 
\colhead{} & 
\colhead{[s]} & 
\colhead{[s]} & 
\colhead{[10$^{21}$ cm$^{-2}$]} &
\colhead{} & 
\colhead{}
}
\startdata
XRF 050406 & 92 & $1.5 \times 10^{5}$ & -- & $-2.3$ & 32.0/16 & $1.1 \times 10^{4}$ & $1.4 \times 10^{6}$ & $3.5_{-2.0}^{+5.9}$ & $-3.5_{-2.3}^{+1.1}$ & 6.3/8\\
XRF 050416A & 85 & $1.4 \times 10^{5}$ & $<11$ & $-2.4_{-1.5}^{+0.8}$ & 9.9/9 & 184 & $6.4 \times 10^{6}$ & $5.6_{-0.9}^{+1.0}$ & $-2.1 \pm 0.1$ & 81.4/100\\
XRF 050714B & 157 & 219 & $7.2_{+1.0}^{+1.2}$ & $-5.8_{-0.6}^{+0.5}$ & 34.2/27 & 257 & $9.5 \times 10^{5}$ & $2.9_{-0.8}^{+1.0}$ & $-2.8_{-0.4}^{+0.3}$ & 21.9/17\\
XRF 050819 & 147 & 202 & $< 0.4$ & $-2.3_{-0.3}^{+0.2}$ & 7.3/10 & 239 & $6.3 \times 10^{5}$ & $< 2$ & $-2.2_{-0.4}^{+0.3}$ & 17.7/11\\
XRF 050824 & -- & -- & -- & -- & -- & $6.2 \times 10^{3}$ & $2.1 \times 10^{6}$ & $2.4_{-0.9}^{+1.0}$ & $-2.2 \pm 0.2$ & 29.4/39\\
XRF 060219 & 126 & $5.7 \times 10^{4}$ & $3.2_{-2.9}^{+6.9}$ & $< -2.6$ & 8.3/9 & 146 & $6.9 \times 10^{5}$ & $3.2_{-0.9}^{+1.1}$ & $-3.1_{-0.5}^{+0.4}$ & 23.4/19\\
XRF 060428B & 212 & 418 & $0.3 \pm 0.1$ & $-2.8 \pm 0.1$ & 126.9/121 & 622 & $1.0 \times 10^{6}$ & $< 0.2$ & $-1.9 \pm 0.1$ & 17.2/30\\
XRF 060512 & 110 & 155 & $0.6_{-0.5}^{+0.7}$ & $-4.4_{-0.7}^{+0.6}$ & 13.9/10 & $3.7 \times 10^{3}$ & $3.5 \times 10^{5}$ & $< 0.3$ & $-1.9_{-0.2}^{+0.1}$ & 24.6/15\\
XRF 060923B & -- & -- & -- & -- & -- & 139 & $6.0 \times 10^{3}$ & $3_{-1}^{+2}$ & $-2.0_{-0.5}^{+0.4}$ & 2.6/8\\
XRF 060926 & 66 & 875 & $25_{-17}^{+28}$ & $-1.9_{-0.3}^{+0.2}$ & 11.4/15 & $4.3 \times 10^{3}$ & $2.8 \times 10^{5}$ & $< 40$ & $-1.8_{-0.4}^{+0.2}$ & 10.0/7\\\hline
XRR 050219B & $3.2 \times 10^{3}$ & $1.2 \times 10^{5}$ & $0.6 \pm 0.2$ & $-1.81_{-0.09}^{+0.08}$ & 152.6/161 & $9.1 \times 10^{3}$ & $3.2 \times 10^{6}$ & $1.0_{-0.5}^{+0.6}$ & $-2.0 \pm 0.2$ & 26.6/22\\
XRR 050410 & $1.9 \times 10^{3}$ & $7.9 \times 10^{4}$ & $13_{-9}^{+18}$ & $< -3.3$ & 28.7/26 & $1.9 \times 10^{3}$ & $9.2 \times 10^{5}$ & $< 8$ & $-1.7_{-1.0}^{+0.5}$ & 23.1/13\\
XRR 050525A & -- & -- & -- & -- & -- & $5.9 \times 10^{3}$ & $3.0 \times 10^{6}$ & $2 \pm 1$ & $-2.1 \pm 0.2$ & 31.8/41\\
XRR 050713A & 80 & $1.2 \times 10^{4}$ & $2.4 \pm 0.3$ & $-2.41_{-0.09}^{+0.08}$ & 146.5/166 & $4.3 \times 10^{3}$ & $1.7 \times 10^{6}$ & $2.5 \pm 0.5$ & $-2.1 \pm 0.1$ & 57.6/78\\
XRR 050815 & -- & -- & -- & -- & -- & 89 & $1.8 \times 10^{5}$ & $< 2$ & $-1.8_{-0.4}^{+0.3}$ & 9.7/11\\
XRR 050915B & 150 & $6.5 \times 10^{4}$ & $< 0.5$ & $-2.6_{-0.2}^{+0.1}$ & 53.7/53 & 288 & $9.6 \times 10^{5}$ & $< 1$ & $-2.2_{-0.3}^{+0.2}$ & 25.7/24\\
XRR 051021B & 86 & 115 & $< 10$ & $-1.2_{-1.1}^{+0.5}$ & 1.6/2 & 258 & $5.2 \times 10^{5}$ & $< 4$ & $-2.0_{-0.4}^{+0.2}$ & 9.1/14\\ 
XRR 060111A & 74 & 517 & $1.7 \pm 0.1$ & $-2.33_{-0.05}^{+0.04}$ & 367.6/300 & $3.8 \times 10^{3}$ & $7.6 \times 10^{5}$ & $1.4_{-0.4}^{+0.5}$ & $-2.2 \pm 0.2$ & 33.7/39\\
XRR 060115 & 121 & $5.4 \times 10^{3}$ & $< 10$ & $-1.84_{-0.09}^{+0.08}$ & 78.9/85 & 616 & $4.6 \times 10^{5}$ & $< 8$ & $-2.3_{-0.2}^{+0.1}$ & 21.6/26\\
XRR 060206 & 64 & $3.7 \times 10^{4}$ & $14_{-7}^{+8}$ & $-2.4_{-0.2}^{+0.1}$ & 72.3/79 & $1.7 \times 10^{3}$ & $3.7 \times 10^{6}$ &
 $12_{-10}^{+11}$ & $-2.0_{-0.2}^{+0.1}$ & 46.5/45\\
XRR 060211A & 172 & 379 & $0.6 \pm 0.2$ & $-1.95 \pm
 0.07$ & 162.1/172 & 662 & $5.7 \times 10^{5}$ & $1.3_{-0.7}^{+0.8}$
 & $-2.1 \pm 0.2$ & 16.2/23\\
XRR 060510A & 98 & 143 & -- & $-3.7$ & 25.5/8 & $2.4 \times 10^{4}$ & $5.7
 \times 10^{5}$ & $< 0.4$ & $-2.03_{-0.10}^{+0.06}$ & 121.1/100\\
XRR 060707 & 127 & 160 & $< 6$ & $-1.8_{-0.3}^{+0.2}$ &
 6.6/5 & 488 & $2.8 \times 10^{6}$ & $10 \pm 7$ & $-2.1 \pm 0.1$ & 33.6/39\\
XRR 060814 & 163 & $5.2 \times 10^{4}$ & $2.6 \pm 0.2$ & $-2.01 \pm
 0.05$ & 363.1/280 & $1.1 \times 10^{3}$ & $1.4 \times 10^{6}$ & $3.1
 \pm 0.3$ & $-2.33 \pm 0.08$ & 169.6/158\\
XRR 060825 & 199 & $1.1 \times 10^{5}$ & $< 8$ &
 $-1.6_{-1.3}^{+0.5}$ & 5.4/4 & 92 & $5.9 \times 10^{5}$ &
 $3_{-2}^{+4}$ & $-1.9 \pm 0.5$ & 8.9/10\\
XRR 060904A & 97 & $2.1 \times 10^{3}$ & $ 1.8_{-0.1}^{+0.2}$ &
 $-2.61_{-0.08}^{+0.07}$ & 255.7/208 & $5.4 \times 10^{4}$ & $1.3 \times
 10^{6}$ & $3_{-1}^{+2}$ & $-2.9_{-0.8}^{+0.5}$ & 10.3/10\\
XRR 060927 & -- & -- & -- & -- & -- & 147 & $2.1 \times 10^{5}$ &
 $< 37$ & $-1.8 \pm 0.2$ & 6.7/12\\\hline
GRB 050124 & -- & -- & -- & -- & -- & $1.1 \times 10^{4}$ & $5.0 \times 10^{6}$ & $<0.8$ & $-1.9_{-0.3}^{+0.2}$ & 13.0/14\\
GRB 050128 & -- & -- & -- & -- & -- & $4.5 \times 10^{3}$ & $9.9 \times 10^{4}$ & $0.7_{-0.2}^{+0.3}$ & $-2.1 \pm 0.1$ & 83.6/82\\
GRB 050219A & 112 & $5.7 \times 10^{3}$ & $1.8_{-0.4}^{+0.5}$ & $-2.1
 \pm 0.2$ & 50.7/55 & 456 & $3.2 \times 10^{6}$ & $< 8$
 & $-1.8_{-1.3}^{+0.5}$ & 6.6/4\\
GRB 050326 & $3.3 \times 10^{3}$ & $9.9 \times 10^{3}$ &
 $0.9_{-0.6}^{+0.7}$ & $-2.0_{-0.3}^{+0.2}$ & 22.3/25 &
 $5.0 \times 10^{3}$ & $5.3 \times 10^{5}$ & $0.6_{-0.5}^{+0.6}$ 
& $-2.0 \pm 0.2$ & 27.3/26\\
GRB 050401 & 133 & $8.5 \times 10^{3}$ & $14 \pm 2$ & $-1.91 \pm
 0.04$ & 277.1/266 & $8.1 \times 10^{3}$ & $1.1 \times 10^{6}$ &
 $21_{-11}^{+17}$ & $-2.0 \pm 0.2$ & 22.9/25\\
GRB 050603 & -- & -- & -- & -- & -- & $3.4 \times 10^{4}$ & $1.8 \times
 10^{6}$ & $6 \pm 4$ & $-1.98_{-0.06}^{+0.12}$ & 29.0/49\\
GRB 050716 & 105 & $7.6 \times 10^{4}$ & $< 0.1$ & $-1.34_{-0.05}^{+0.03}$ 
& 208.9/202 & $4.1 \times 10^{3}$ & $1.8 \times 10^{6}$ & $0.6 \pm 0.5$ 
 & $-2.1 \pm 0.2$ & 43.3/36\\
GRB 050717 & 91 & $2.7 \times 10^{4}$ & $1.8_{-0.6}^{+0.7}$ & $-1.5
 \pm 0.1$ & 110.7/105 & 4000 & $6.0 \times 10^{5}$ & $< 2$
 & $-1.5_{-0.3}^{+0.2}$ & 23.6/15\\
GRB 050922C & 116 & $6.2 \times 10^{4}$ & $< 2$ & $-2.02 \pm
 0.07$ & 107.9/124 & 3998 & $5.9 \times 10^{5}$ & $7 \pm 3$ & $-2.53_{-0.08}^{+0.07}$ & 60.2/49\\
GRB 051109A & 128 & $1.7 \times 10^{4}$ & $< 4$ & $-2.0 \pm
 0.1$ & 42.9/32 & $3.4 \times 10^{3}$ & $1.5 \times 10^{6}$ & $5 \pm
 3$ & $-2.08 \pm 0.07$ & 130.7/129\\
GRB 060105 & 97 & $4.6 \times 10^{3}$ & $1.6 \pm 0.1$ & $-1.99 \pm
 0.03$ & 527.6/496 & $1.0 \times 10^{4}$ & $3.8 \times 10^{5}$ & $1.7
 \pm 0.4$ & $-2.2 \pm 0.1$ & 84.8/94\\
GRB 060204B & 103 & $1.8 \times 10^{4}$ & $1.9 \pm 0.2$ & $-2.28_{-0.09}^{+0.08}$ 
& 122.5/129 & $4.0 \times 10^{3}$ & $8.1 \times 10^{5}$ & 
$1.3 \pm 0.3$ & $-2.3_{-0.2}^{+0.1}$ & 54.9/56\\ 
GRB 060813 & 85 & $7.6 \times 10^{4}$ & $1.1 \pm 0.4$ & $-1.88 \pm
 0.08$ & 167.1/163 & $4.1 \times 10^{3}$ & $2.6 \times 10^{5}$ & $1.3
 \pm 0.4$ & $-2.0 \pm 0.1$ & 105.1/102\\
GRB 060908 & 80 & $1.3 \times 10^{4}$ & $< 8$ & $-2.3 \pm 0.2$ & 18.9/26 & $1.2 \times 10^{3}$ & $1.1 \times 10^{6}$ & $< 11$ & $-2.0_{-0.3}^{+0.2}$ & 13.7/14\\
\enddata
\tablenotetext{\dagger}{The definition of the photon index, $\Gamma$, is based on the spectral model: f(E) = KE$^{\Gamma}$.}
\end{deluxetable}

%% file: tab4.tex
\begin{deluxetable}{lccccccccc}
\tabletypesize{\scriptsize}
\rotate
\tablecaption{XRT X-ray temporal properties of 41 Swift Bursts\label{tbl:xray_temp}}
\tablewidth{0pt}
\tablehead{
\colhead{GRB} & 
\colhead{$t_{min}$} & 
\colhead{$t_{max}$} & 
\colhead{$\alpha^{ini}_{1}(\star)$} & 
\colhead{$t_{br}^{ini}(\diamond)$} &
\colhead{$\alpha^{ini}_{2}(\dagger)$} & 
\colhead{$\alpha^{fin}_{2,3}(\ast)$} & 
\colhead{$t_{br}^{fin}(\circ)$} & 
\colhead{$\alpha^{fin}_{3,4}(\bullet)$} & 
\colhead{$\chi^{2}$/d.o.f.}\\
\colhead{} & 
\colhead{[s]} & 
\colhead{[s]} & 
\colhead{} & 
\colhead{[s]} & 
\colhead{} & 
\colhead{} & 
\colhead{[s]} & 
\colhead{} & 
\colhead{}}
\startdata
XRF 050406  & 170 & $1.1 \times 10^{6}$ & $-1.7 \pm 0.3$ & 4360& $-0.7 \pm 0.2$ & -- & -- & -- & 3.8/4\\
XRF 050416A & 90 & $6.1 \times 10^{6}$ & $-1.6 \pm 0.8$ & $200 \pm 90$ & $-0.63 \pm 0.04$ & -- & $7700 \pm 110$ & $-0.86 \pm 0.03$ & 86.2/79\\
XRF 050714B & 163 & $8.3 \times 10^{5}$ & $-7.2 \pm 0.7$ & $270 \pm 20$ & $-1.1 \pm 1.5$ & -- & 4100 & $-0.70 \pm 0.07$ & 27.1/15\\
XRF 050819  & 154 & $4.6 \times 10^{5}$ & $-2.2 \pm 2.0$ & $190 \pm 20$ & $-5.6 \pm 1.0$ & $0.07 \pm 0.50$ & $(2.2 \pm 0.1) \times 10^{5}$ & $-1.2 \pm 0.6$ & 4.4/5\\
XRF 050824  & 6550 & $2.0 \times 10^{6}$ & $-0.4 \pm 0.1$ & $(5.9 \pm 2.4) \times 10^{4}$ & $-0.87 \pm 0.09$ & -- & -- & -- & 36.6/24\\
XRF 060219  & 129 & $3.9 \times 10^{5}$ & $-4.8 \pm 0.8$ & $280 \pm 40$ & $-0.4 \pm 0.1$ & -- & $1700 \pm 50$ & $-1.1 \pm 0.1$ & 8.5/7\\
XRF 060428B & 212 & $6.4 \times 10^{5}$ & $-4.4 \pm 0.1$ & $670 \pm 30$ & $-0.98 \pm 0.04$ & -- & -- & -- & 70.3/61\\
XRF 060512  & 115 & $2.9 \times 10^{5}$ & $-1.30 \pm 0.03$ & -- & -- & -- & -- & -- & 14.1/13\\
XRF 060923B & 145 & 5820 & $-0.60 \pm 0.07$ & -- & -- & -- & -- & -- & 6.6/8\\
XRF 060926  & 192 & $2.0 \times 10^{5}$ & $-0.2 \pm 0.2$ & $1500 \pm 550$ & $-1.4 \pm 0.2$ & -- & -- & -- & 7.8/8\\\hline
XRR 050219B & 3200 & $3.1 \times 10^{6}$ & $-1.29 \pm 0.04$ & -- & -- & -- & -- & -- & 37.1/28\\
XRR 050410  & 246 & $6.1 \times 10^{5}$ & $-0.98 \pm 0.09$  & -- & -- & -- & -- & -- & 6.4/4\\
XRR 050525A & 77 & $2.4 \times 10^{6}$ & $-0.8 \pm 0.2$ & 2590 & $-1.53 \pm 0.06$ & -- & -- & -- & 29.6/27\\
XRR 050713A & 330 & $9.2 \times 10^{4}$ & $-0.18 \pm 2.00$ & 1450 & $-1.05 \pm 0.07$ & -- & $(4.5 \pm 0.1) \times 10^{4}$ & $-1.9 \pm 0.4$ & 36.6/29 \\
XRR 050815  & 3550 & $1.5 \times 10^{5}$ & $-1.9 \pm 0.3$ & -- & -- & -- & -- & -- & 11.3/9\\
XRR 050915B & 151 & $9.1 \times 10^{5}$ & $-5.5 \pm 0.4$ & $350 \pm 35$ & $-1.6 \pm 0.4$ & $-0.4 \pm 0.2$ & $(3.4 \pm 0.2) \times 10^{4}$ & $-0.9 \pm 0.2$ & 33.3/19\\
XRR 051021B & 98 & $4.8 \times 10^{5}$ & $-1.9 \pm 0.1$ & $3310 \pm 1720$ & $-0.6 \pm 0.2$ & -- & -- & -- & 15.4/14\\
XRR 060111A & 3810 & $7.5 \times 10^{5}$ & $-0.90 \pm 0.04$ & -- & -- & -- & -- & -- & 27.3/27\\
XRR 060115  & 122 & $3.6 \times 10^{5}$ & $-4.4 \pm 0.8$ & $150 \pm 10$ & $-2.4 \pm 0.2$ & $-0.6 \pm 0.1$ & $(3.1 \pm 1.7) \times 10^{4}$ & $-1.2 \pm 0.2$ & 29.4/30\\
XRR 060206  & 69 & $9.1 \times 10^{5}$ & $-0.4 \pm 0.2$ & -- & -- & -- & $1.1 \times 10^{4}$ & $-1.3 \pm 0.1$ & 21.0/12\\
XRR 060211A & 186 & $5.3 \times 10^{5}$ & $-7 \pm 3$ & --  & -- & $-2.1 \pm 0.4$ & $2960 \pm 1110$ & $-0.60 \pm 0.09$ & 72.1/63\\
XRR 060510A & 105 & $5.3 \times 10^{5}$ & $-3 \pm 1$ & $130 \pm 8$ & $-0.01 \pm 0.06$ & -- & $5500 \pm 640$ & $-1.48 \pm 0.04$ & 161.8/156\\
XRR 060707  & 207 & $2.0 \times 10^{6}$ & $-2.5 \pm 0.3$ & $640 \pm 100$ & $-0.4 \pm 0.1$ & -- & $(2.9 \pm 1.6) \times 10^{4}$ & $-1.1 \pm 0.1$ & 34.0/25\\
XRR 060814  & 78 & $1.1 \times 10^{6}$ & $-2.56 \pm 0.05$ & $940 \pm 55$ & $-0.29 \pm 0.05$ & $-1.06 \pm 0.06$ & $(4.2 \pm 1.7) \times 10^{4}$ & $-1.39 \pm 0.07$ & 197.9/185\\
XRR 060825 & 108 & $4.1 \times 10^{5}$ & $-0.96 \pm 0.04$ & -- & -- & -- & -- & -- & 10.7/7\\
XRR 060904A & 83 & $1.2 \times 10^{6}$ & $-3.6 \pm 0.2$ & $190 \pm 80$ & $-1.1 \pm 0.1$ & -- & -- & -- & 40.1/39\\
XRR 060927 & 93 & $1.3 \times 10^{5}$ & $-0.71 \pm 0.06$ & $4400 \pm 60$ & $-1.9 \pm 0.5$ & -- & -- & -- & 8.1/13\\\hline
GRB 050124 & $1.1 \times 10^{4}$ & $7.9 \times 10^{4}$ & $-1.6 \pm 0.1$ & -- & -- & -- & -- & -- & 14.0/11\\
GRB 050128 & 240 & $7.9 \times 10^{4}$ & $-0.8 \pm 0.1$ & $1700 \pm 570$ & $-1.24 \pm 0.04$ & -- & -- & -- & 138.7/128\\
GRB 050219A & 116 & $2.4 \times 10^{4}$ & $-3.2 \pm 0.3$ & $256 \pm 20$ & $-0.95 \pm 0.07$ & -- & -- & -- & 28.0/11\\
GRB 050326 & 3350 & $1.9 \times 10^{5}$ & $-1.70 \pm 0.05$ & -- & -- & -- & -- & -- & 15.6/21\\
GRB 050401 & 136 & $7.2 \times 10^{5}$ & $-0.64 \pm 0.04$ & $840 \pm 60$ & $-0.47 \pm 0.05$ & -- & $3440 \pm 630$ & $-1.4 \pm 0.1$ & 126.6/115\\
GRB 050603 & $3.7 \times 10^{4}$ & $1.4 \times 10^{6}$ & $-1.7 \pm 0.1$ & -- & -- & -- & -- & -- & 28.5/34\\
GRB 050716 & 535 & $1.6 \times 10^{6}$ & $-1.10 \pm 0.04$ & -- & -- & -- & -- & -- & 31.8/34\\
GRB 050717 & 93 & $5.4 \times 10^{5}$ & $-2.0 \pm 0.1$ & $320 \pm 70$ & $-1.36 \pm 0.05$ & -- & -- & -- & 49.8/49\\
GRB 050922C & 119 & $5.4 \times 10^{5}$ & $-0.8 \pm 0.2$ & $280 \pm 70$ & $-1.18 \pm 0.03$ & -- & $(2.1 \pm 0.7) \times 10^{4}$ & $-1.8 \pm 0.2$ & 188.6/78\\
GRB 051109A & 137 & $1.5 \times 10^{6}$ & $-2.9 \pm 0.2$ & $2670 \pm 300$ & $-0.9 \pm 0.2$ & $-1.12 \pm 0.07$ & $(5.2 \pm 1.0) \times 10^{4}$ & $-1.38 \pm 0.06$ & 179.5/152\\
GRB 060105  & 97 & $3.7 \times 10^{5}$ & $-1.18 \pm 0.06$ & $199 \pm 2$ & $0.78 \pm 0.02$ & $-0.5 \pm 0.2$ & $(5.72 \pm 0.03) \times 10^{4}$ & $-2.1 \pm 0.2$ & 501.5/438\\
GRB 060204B & 450 & $4.9 \times 10^{5}$ & $-0.68 \pm 0.08$ & $5170 \pm 190$ & $-0.02 \pm 0.80$ & -- & $6670 \pm 910$ & $-1.51 \pm 0.09$ & 44.4/48\\
GRB 060813  & 115 & $1.9 \times 10^{5}$ & $-0.66 \pm 0.05$ & $1680 \pm 380$ & $-1.22 \pm 0.04$ & -- & $(5.0 \pm 0.5) \times 10^{4}$ & $-2.6 \pm 0.4$ & 150.5/146\\
GRB 060908 & 85 & $9.5 \times 10^{5}$ & $-0.68 \pm 0.06$ & $875 \pm 1$ & $-1.62 \pm 0.09$ & -- & $(1.3 \pm 0.7) \times 10^{4}$ & $-0.8 \pm 0.1$ & 51.7/51\\
\enddata
\tablenotetext{\star}{The decay index of the first power-law component.  
For most of cases, this component corresponds to the very steep decay 
$\alpha_{1}$ as discussed in $\S$1.}
\tablenotetext{\diamond}{The break time of the first component in seconds 
after the BAT trigger.}
\tablenotetext{\dagger}{The post break decay power-law index of the first component.  
For most cases, this component corresponds to the shallow decay $\alpha_{2}$ as discussed in $\S$1.}
\tablenotetext{\ast}{The pre-break decay index of the last component.}
\tablenotetext{\circ}{The break time of the last component in seconds after the BAT 
trigger.  For most cases, this component corresponds to either the shallow decay 
$\alpha_{2}$ or the steeper decay $\alpha_{3}$, as discussed in $\S$1.}
\tablenotetext{\bullet}{The post break decay power-law index of the last component.  
For most cases, this component corresponds to either the steeper decay 
$\alpha_{3}$ or the much steeper decay $\alpha_{4}$, as discussed in $\S$1.}
\end{deluxetable}